\documentclass[%
reprint,
showpacs,
amsmath,amssymb, aps,
prb,
]{revtex4-1}
\usepackage{amsmath}
\usepackage{graphicx}
\usepackage{dcolumn}
\usepackage{bm}
\usepackage{hyperref}

\hypersetup{colorlinks,linkcolor={red},citecolor={blue},urlcolor={blue}}

\def\be{\begin{equation}}
\def\ee{\end{equation}}
\def\bea{\begin{eqnarray}}
\def\eea{\end{eqnarray}}
\def\c{{\bf c}}

\def\d{{\bf d}}
\def\v{{\bf v}}
\def\kv{{\bf k}}

\def\Sv{{\bf S}}

\def\sv{{\bm \sigma}}
\def\Rv{{\bf R}}
\def\rv{{\bf r}}

\def\la{\langle}

\def\t{\tau}
\def\la{\lambda}
\def\ve{\varepsilon}
\def\f{{\rm F}}
\def\r{{\rm R}}

\def\tver{\tilde{\varepsilon}_{\rm R}}
\usepackage{color}

\newcommand{\ket}[1]{\left|{#1}\right\rangle}

\begin{document}

\title{Transport in two dimensional Rashba electron systems doped with interacting magnetic impurities}

\author{A. N. Zarezad}
\affiliation{Department of Physics, Institute for Advanced Studies in Basic Sciences (IASBS), Zanjan 45137-66731, Iran}
\author{J. Abouie}
\email[]{jahan@iasbs.ac.ir}
\affiliation{Department of Physics, Institute for Advanced Studies in Basic Sciences (IASBS), Zanjan 45137-66731, Iran}

\begin{abstract}
	We study the transport properties of two dimensional electron systems with strong Rashba spin-orbit coupling (SOC) doped with interacting magnetic impurities.  
	Interactions between magnetic impurities cause the formation of magnetic clusters with temperature dependent mean sizes (CMSs) distributed randomly on the surface of the system. Treating magnetic clusters as scattering centers, by employing a generalized relaxation time approximation we obtain the non-equilibrium distribution functions of Rashba electrons in both regimes of above and below the band-crossing point (BCP) and present the explicit forms of the conductivity in terms of effective relaxation times. We demonstrate that the combined effects of SOC and magnetic clusters cause the system to be anisotropic and the magneto-resistance strongly depends on both the clusters' mean size and spin, the strengths of SOC and the location of Fermi energy with respect to the BCP. Our results show that there are many contrasts between the transport properties of the system in the two regimes of above and below the BCP. 
	By comparing the anisotropic magneto-resistance (AMR) of the two dimensional Rashba systems with the surface AMR of three dimensional magnetic topological insulators, we also point out the differences between these systems.
\end{abstract}

\date{\today}


\maketitle


\section{Introduction}\label{sec:intro}

Two dimensional electron systems with Rashba spin-orbit coupling (SOC) play leading roles in novel spintronics. Locking of the spin degrees of freedom with the spatial motion of itinerant electrons splits the spin-degenerate band of the system into two parabolic bands with opposite spin-helicity intersecting each other at the band-crossing point (BCP). The presence of two bands with different spin states is one of the hallmarks of two dimensional Rashba electron systems (2DRSs) in spintronics applications, especially in the manipulation of spin state of electrons in the absence of external magnetic fields.\cite{rashba,jetp1,jetp2,Sharma531,RevModPhys.76.323}
Topological changes of Fermi surface and the variations of band structure from two spin states to the valley band with a single spin state in passing through the BCP, lead to several phenomena in low-density regime, \cite{PhysRevLett.98.167002,PhysRevB.72.075307,doi:10.1063/1.4902134} 
such as qualitatively significant modifications of thermoelectric and thermopower properties,\cite{PhysRevB.93.075150,doi:10.1063/1.4902134} modifications of the classical Dyakonov-Perel mechanism of spin relaxation,\cite{PhysRevB.72.075307} and the enhancement of the superconducting critical temperature in 2DRSs with strong SOC.\cite{PhysRevLett.98.167002}

Doping of 2DRSs with magnetic impurities leads to many exotic phenomena\cite{PhysRevLett.100.236602,PhysRevB.84.193411,PhysRevLett.108.046601,PhysRevB.85.081107,Yan2012} such as anomalous Hall effect and anisotropic magnetoresistance (AMR). The combined effects of Rashba SOC and localized spins significantly modify the transport properties of the system, they lead to resonantly enhancement of anomalous Hall conductivity\cite{PhysRevLett.100.236602} and giant AMR.\cite{PhysRevB.79.045427}  
For systems with low impurity concentrations, interactions between magnetic impurities are negligible and the transport properties of the system is properly given by considering single impurity scattering effects,\cite{PhysRevB.79.045427} however when doping increases these interactions become crucial.
One of the effective solution for systems with high impurity concentrations is taking the effects of multiple scatterings into account.\cite{PhysRevB.73.035325} 
In this paper we use the concept of magnetic clusters and study the effects of magnetic clusters on the transport properties of 2DRSs. Exchange interactions between magnetic impurities cause the formation of clusters where constructed of correlated magnetic impurities fluctuating consistently in the same direction. The clusters' mean sizes (CMSs) and their number (CN) depend on temperature. 
The cluster model was first proposed in Refs.[\onlinecite{PhysRevB.84.024428,PhysRevB.77.165433}] for explaining the temperature dependence of the magneto-resistance of magnetic semiconductors, and recently developed for investigating the surface conductivity of three dimensional magnetic topological insulators.\cite{PhysRevB.98.155413}
In this paper by treating magnetic clusters as scattering centers, distributed randomly on the surface of the 2DRSs, and modeling the interaction of itinerant electrons with magnetic clusters by a long-range scattering potential, we demonstrate that the combined effects of Rashba SOC and magnetic clusters cause the system to be anisotropic and the anisotropy strongly depends on both the clusters' mean size and spin, the strengths of SOC and the locations of Fermi energy with respect to the BCP. 
We obtain the non-equilibrium distribution functions of electrons in different bands within the semiclassical Boltzmann approach, and compute the conductivity and the AMR of the system in both regimes of above and below the BCP. 
We demonstrate that for large CMSs the angular dependence of the AMR is unconventional in comparison with conventional ferromagnets.
We also show that, there are many contrasts between the transport properties of the system in the two regimes of above and below the BCP. 

This paper is organized as follows. In Sec. \ref{sec:model}, we obtain the band structure of 2DRSs and present the generalized relaxation time approximation in order to find the non-equilibrium distribution functions of electrons in each bands. In Sec. \ref{sec:two-band}, we present the explicit form of the conductivities in the two-band regime. In this section we also compare the relaxation times obtained by other approaches. The results of the single band regime are presented in Sec. \ref{sec:single-band}. The summaries and conclusions are given in Sec.\ref{sec:conclusion}. At the end of this paper we present the details of our calculations in three different appendices.

\section{Scattering potential and GRTA}\label{sec:model}

\begin{figure}[h]
	\centerline{\includegraphics[width=70mm]{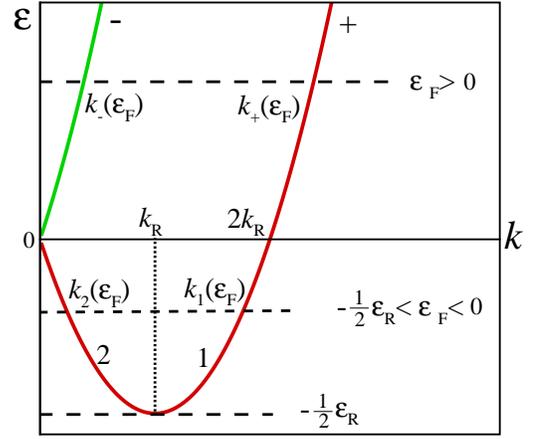}}
	\caption{(Color online)  The band structure of the 2DRS. 
		Above the BCP ($\ve_\f> 0$) Fermi energy intersects the band $n$, at the point:  $k_n(\ve_\f)=k_\f^0(n\sqrt{\tver}+\sqrt{1+\tver})$. 
		However, when the Fermi energy lies in the interval  $-\frac{\ve_\r}{2}<\ve_\f<0$, it intersects the branch $\nu$ at the point $k_\nu(\ve_\f)=k_\f^0(\sqrt{\tver}-(-1)^\nu\sqrt{\tver-1})$. The parameter $k_\f^0=\sqrt{2m\ve_\f/\hbar^2}$ is the Fermi wave number of free electrons (without SOC). $\tver$ is defined as $\ve_\r/2\ve_\f$ for $\ve_\f>0$, and $\ve_\r/(2|\ve_\f|)$ for $\ve_\f<0$.
	} \label{fig:band}
\end{figure}

The Hamiltonian of a 2DRS is given by:
\begin{equation}
H_0=\frac{\hbar^{2}k^2}{2m}+\alpha(\sigma_yk_x-\sigma_xk_y),
\label{eq:hamiltonian}
\end{equation}
where the first term is the kinetic energy of electrons with effective mass $m$, and the second term is the Rashba SOC with the strength of $\alpha$.  The Pauli matrices $\sigma_{x,y}$ indicate spin of electrons, and $k_{x,y}$ are the two components of electrons' wave vector, $\kv$. The eigenenergies and eigenstates of
the Hamiltonian $H_0$ are obtained as,
\begin{equation}
\begin{aligned}
\varepsilon_{n}(k)=\frac{\hbar^{2} k^{2}}{2m}-n k \alpha&
=\frac{\hbar^{2}}{2m}(k-n k_\r)^{2}-\frac{\ve_\r}{2},
\end{aligned}
\label{rashba:spectrum}
\end{equation}
\begin{equation}
\psi_{n}(\kv,\rv)=\frac{e^{i\kv\cdot\rv}}{\sqrt{2A}}\begin{pmatrix}
-n i {e}^{-i \phi} \\
1
\end{pmatrix},
\label{eq:eigenvector}
\end{equation}
where $n (=\pm)$ represents the Rashba bands, $\ve_\r =\frac{m\alpha^{2}}{\hbar^{2}}$ is the Rashba energy ($\ve_\r/2$ is the minimum value of the band $+$ at $k_\r=\frac{m\alpha}{\hbar^{2}}$), $A$ is the area of the 2DRS, and $\phi=\arctan(k_y/k_x)$ is the polar angle of the $ \kv$-vector.
In the presence of SOC, the energy spectrum splits into two energy bands + and $-$, intersecting each other at the band-crossing point (BCP), illustrated in Fig. \ref{fig:band}.
Strong SOC has been observed in several materials,\cite{PhysRevLett.98.186807,PhysRevB.77.081407,PhysRevB.80.035438,PhysRevLett.103.046803} such as ${\rm Te}$-terminated surface of the polar semiconductor ${\rm BiTeX}$ (X=I, Br and Cl)\cite{ishizaka,PhysRevLett.108.246802,Eremeev_2013,PhysRevLett.109.116403,Landolt_2013,PhysRevLett.109.096803} where Rashba SOC is at the order of 1.7-3.8 eV$\mathring{A}$, which is one order of magnitude larger than the Rashba SOC in conventional III-V semiconductor heterostructures. 

\begin{figure}[h]
    \centerline{\includegraphics[width=70mm]{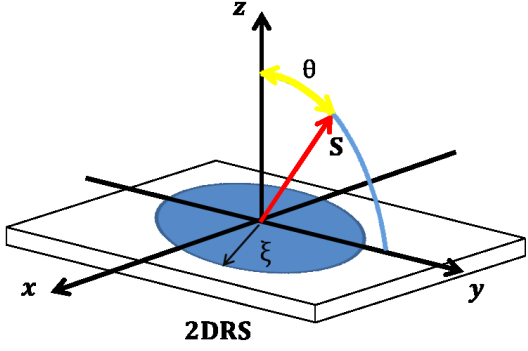}}
   \caption{(Color online) The schematic illustration of a single magnetic cluster with size of $\xi$, centered at ${\bf R}=0$. Red arrow represents cluster's spin ${\bf S}$, and $\theta$ is the tilt angle of $\Sv$ with the $z$ axis. $\Sv$ is subjected to rotate in the $zy$ plane.} \label{fig:schematic}
      \end{figure}

In a two-dimensional electron system doped with magnetic impurities, the interaction between an electron with spin $\sv$ at the position $\rv$, with an impurity with spin $\Sv$ located at $\Rv$ is given by the following Hamiltonian:
    \begin{equation}
    H_{\sigma S}=-J(\rv-\Rv)\sv(\rv)\cdot\Sv(\Rv),
    \end{equation}
where $J(\rv-\Rv)$ is the exchange coupling.  
In dilute magnetic systems itinerant electrons interact with individual single magnetic impurities, and the exchange coupling is modeled by Dirac $\delta$ function as $J(\rv-\Rv)\propto J_0\delta(\rv-\Rv)$, where $J_0$ is a coupling constant at the order of few $meV$.\cite{PhysRevB.78.212405,Liu2006,Semi} When doping of magnetic impurities increases, the interactions between magnetic impurities become significant and we should consider their effects on the transport properties of the system. 
Actually, the exchange interactions between impurities lead to the formation of various magnetic domains with different sizes in the entire system. These ordered domains which are called "magnetic clusters", are constructed of correlated magnetic impurities with the same spin directions. Scattering of electrons by these clusters (rather than single impurities) alters the transport properties of the system.\cite{PhysRevB.98.155413} In order to investigate the effects of magnetic clusters on the conductivity, we model the scattering potential of electrons by a clusters as:
\begin{equation}
H_{\sigma S}=J_0\exp(-|\rv-\Rv|/\xi)\sv(\rv)\cdot\Sv(\Rv),
\label{eq:scattering-Hamiltonian}
\end{equation}
where $\xi$ is clusters' mean size, depending on both temperature and impurity-impurity exchange coupling, and $\sv$ is the spin of magnetic cluster located at $\Rv$.
The scattering potential in Eq. (\ref{eq:scattering-Hamiltonian}) is
long-range and $\xi$ appears as a characteristic length, indicating the range of scattering potential. Without loss of generality, we consider the spins of clusters as classical vectors in $yz$ plane, i.e. $\Sv=S(0,\sin\theta,\cos\theta)$, 
where $\theta$ is the tilt angle of $\Sv$ with the axis normal to the surface of the 2DRS (see Fig. \ref{fig:schematic}). 

In order to obtain the non-equilibrium distribution function of electrons in the presence of magnetic clusters, we use the semi-classical Boltzmann approach.
In the presence of a uniform electric field ${\bf E}$, the non-equilibrium distribution function of electrons in the band $n$,
$f_n(\kv,{\bf E})$, satisfies the following relation:
\begin{equation}
\left(\frac{\partial f_{n}}{\partial t}\right)_{coll}=e{\bf v}_{n}(\kv)\cdot{\bf
E}\left(\frac{\partial
f_n^0}{\partial\varepsilon_n}\right),\label{eq:boltzmann1}
\end{equation}
where ${\bf v}_{n}(\kv)$ is the band velocity, and $f^0_n$ is the Fermi-Dirac distribution
function in the band $n$. By considering elastic scatterings and using detailed balance, 
the left hand side of Eq. (\ref{eq:boltzmann1}) is also written as \cite{PhysRevB.68.165311}:
\begin{equation}
  \begin{aligned}
 \left(\frac{\partial f_{n}}{\partial t}\right)_{coll}=& \sum_{n',\kv'}w_{n,n'}(\kv,\kv')[1-f_{n}(\kv)]f_{n'}(\kv')\\
 & -\sum_{n',\kv'}w_{n',n}(\kv',\kv)[1-f_{n'}(\kv')]f_{n}(\kv),
 \end{aligned}
  \label{eq:collision}
   \end{equation}
where $w_{n,n'}(\kv,\kv')$ is the transition rate between the two eigenstates of the Hamiltonian $H_0$ ($\ket{n \kv}$ and $\ket{n'\kv'}$). 
Using Eqs. (\ref{eq:boltzmann1}) and (\ref{eq:collision}), the Boltzmann equation is written as:
\begin{equation}
\begin{aligned}
 e{\bf v}_{n}(\kv)\cdot{\bf E}\left(\frac{\partial f_n^0}{\partial\varepsilon_n}\right)= \sum_{n',\kv'}w_{n,n'}(\kv,\kv')[f_{n'}(\kv')-f_{n}(\kv)],
\end{aligned}
\label{eq:boltzmann2}
\end{equation}
where the interaction between electrons has been neglected.

In the isotropic 2DRSs ($\theta=0$), the scattering rate depends on the angle between $\kv$ and $\kv'$ ($\Delta\phi=\phi-\phi'$), and relaxation times of electrons in the bands + and $-$ depend only on the magnitudes of $\kv$ and $\kv'$. By using standard methods such as modified relaxation time approximation (MRTA)\cite{PhysRevB.79.045427,PhysRevB.80.134405}, and other analytical exact solutions of self-consistent equations for relaxation times\cite{Xiao2016,PhysRevB.93.075150}, we can solve the equation (\ref{eq:boltzmann2}). But in anisotropic case, when $\theta\neq 0$, the scattering rates as well as the
relaxation times depend on both the magnitudes and the directions of $\kv$
and $\kv'$, so it is no longer possible to use the standard methods for obtaining the non-equilibrium distribution function $f_n$. Using the method developed by Vybrony et. al\cite{PhysRevB.79.045427}, we approximate $f_n$ as:
\begin{equation}
f_{n}-f^0_{n}=e E v_{n}(\kv)\left(\frac{\partial f^0_{n}}{\partial
\ve_n}\right)[a_{n}(\Phi)\cos\chi+ b_{n}(\Phi)\sin\chi],\label{eq:relax-time-anisotropic}
\end{equation}
where $\Phi$ is the polar angle of the band velocity $\v_n$, and $\chi$ is the angle between electric field and $x$ axis. 

According to the band structure, above the BCP band velocities, given by $\v_\pm=\frac{\hbar}{m}(N_0/N_\pm)\kv_\pm$ ($N_\pm(\ve)$ are the density of states (DOS) in the bands $\pm$), are always in the direction of the $\kv$-vector and the polar angle $\Phi$ is equal to $\phi$. The DOS are given by: 
\begin{equation}
N_\pm(\ve)=N_0 \frac{k_\pm(\ve)}{k_\pm(\ve)-n k_\r},
\end{equation}
where $N_0=\frac{m}{2\pi \hbar^{2}}$ is the DOS of 2D free electron systems.
Below the BCP, the energy + behaves non-monotonically, it decreases by increasing $k$ for $k<k_\r$ (branch 2), becomes minimum at $k_\r$, and increases by $k$ for $k>k_\r$ (branch 1). The band velocities for the two branches 1 and 2 are given by $\v_\nu=(-1)^\nu\frac{\hbar}{m}(N_0/N_\nu)\kv_\nu$, where the DOS $N_\nu(\ve)$ is given by: 
\begin{equation}
N_\nu(\ve)=N_0 \frac{k_\nu(\ve)}{|k_\nu(\ve)- k_\r|}.
\end{equation}
In the branch 1, the band velocity $\v_1$ and the $\kv$-vector are parallel and $\Phi=\phi$. For this branch we will write the non-equilibrium distribution function the same as the two-band case. However, in the branch 2 the band velocity $\v_2$ is anti-parallel with $\kv$, and $\Phi=-\phi$. The distribution functions of electrons in this branch will be expressed in terms of the polar angle of $\kv$-vector, with some modifications.

Following we investigate the transport properties of the 2DRSs, for the two regimes of $\ve_\f>0$ and $\ve_\f<0$, in the two separate sections. 

\section{Two-band scattering ($\ve_\f>0$)}\label{sec:two-band}

The conductivity of electron systems is obtained from the following general formula:
   \begin{equation}
   \sigma_{ij}=\frac{-e}{E_j}\sum_{n}\int \frac{d^2k}{(2\pi)^2}v^i_n(\kv) f_n(\kv,\bf E),
   \label{conductivity}
   \end{equation}
where $i$ and $j$ denote $x$ and $y$ directions, and $v^i_n(\kv)$ is the band velocity along $i$ direction. Above the BCP the band velocity is given by $\v_n(\kv)=v_n(k)(\cos\phi,\sin\phi)$ and the conductivities in $x$ and $y$ directions are obtained as (see appendix \ref{Appendix-two-band}):
     \begin{equation}
     \begin{aligned}
      & \sigma_{xx}=\frac{e^2}{4\pi}\int k dk \sum_{n=\pm}\big[v_n^2(k)\left(-\frac{\partial f^0_n}{\partial\ve_n}\right)c^{n}_1(k)\big],  \label{eq:sigma-xx-yy}\\
            & \sigma_{yy}=\frac{e^2}{4\pi}\int k dk \sum_{n=\pm}\big[v_n^2(k)\left(-\frac{\partial f^0_n}{\partial\ve_n}\right)s^{n}_1(k)\big],
              \end{aligned}
            \end{equation}
where the coefficients $c_{1 }^{\pm}(k)$ and $s_{1 }^{\pm}(k)$, introduced in the  appendix \ref{Appendix-two-band}, have a dimension of time. They depend on the parameters $k$, $\theta$, $\xi$, and $\alpha$. We threat them as momentum relaxation times of electrons along $x$ and $y$ directions and define the following dimensionless variables: $\tau_{ x}^\pm=\omega_0 c_{1}^{\pm}$ and $\tau_{y}^\pm=\omega_0 s_{1}^{\pm}$, where $\omega_{0}=\frac{\pi \hbar \eta_c J_{0}^{2}S^2 }{4 A m\varepsilon_\f}$ is a scale factor with units of scattering rate. 

Since the temperature region where $\xi$ and $\eta_{c}$ vary is much smaller than the Fermi temperature of the system, we can approximate the function $\partial f^0_n/\partial\ve_n$ with the Dirac delta function $\delta(\ve-\ve_\f)$. For example the Curie temperature for the ferromagnetic IV-VI compounds like Ge$_{1-x}$Mn$_x$Te is about $80$ K.\cite{doi:10.1063/1.1555697} Moreover, in a large class of materials containing heavy $5d$ elements together with rare-earth or transition metal elements, the critical temperature is very lower than the Fermi temperature. These materials are non-magnetic in the bulk but exhibit 2D magnetism at the surface. For example EuIr$_{2}$Si$_2$ with nonmagnetic bulk, reveals controllable 2D ferromagnetism below 48 K.\cite{7591d715a18e40f28a1d62e4885a4bd6} 
       
With the above assumptions, the conductivities (\ref{eq:sigma-xx-yy}) reduce to:  
       \begin{equation}
       \begin{aligned}
       &\sigma          _{xx}=\sigma_{0}\sqrt{\tver+1}\sum_{n=\pm}\tau^{n}_x(\ve_\f)\Big[n\sqrt{\tver}+\sqrt{\tver+1}\Big],\\
       &\sigma          _{yy}=\sigma_{0}\sqrt{\tver+1}\sum_{n=\pm}\tau^{n}_y(\ve_\f)\Big[n\sqrt{\tver}+\sqrt{\tver+1}\Big], \label{sigma-two-band-iso}
       \end{aligned}
       \end{equation}
       where $\sigma_{0}=(\frac{e^{2}}{2 h})\frac{8m A \ve_\f^3}{\pi \eta_c J_{0}^{2}S^2\hbar^2}$ is a scale factor with units of conductivity. It is actually the conductivity of the system in the absence of SOC when $\xi$ is about the Fermi wavelength of free electrons $\lambda_\f^0=h/
       \sqrt{2m\ve_\f}$. 

\begin{figure}[h]
   	\centerline{\includegraphics[width=45mm]{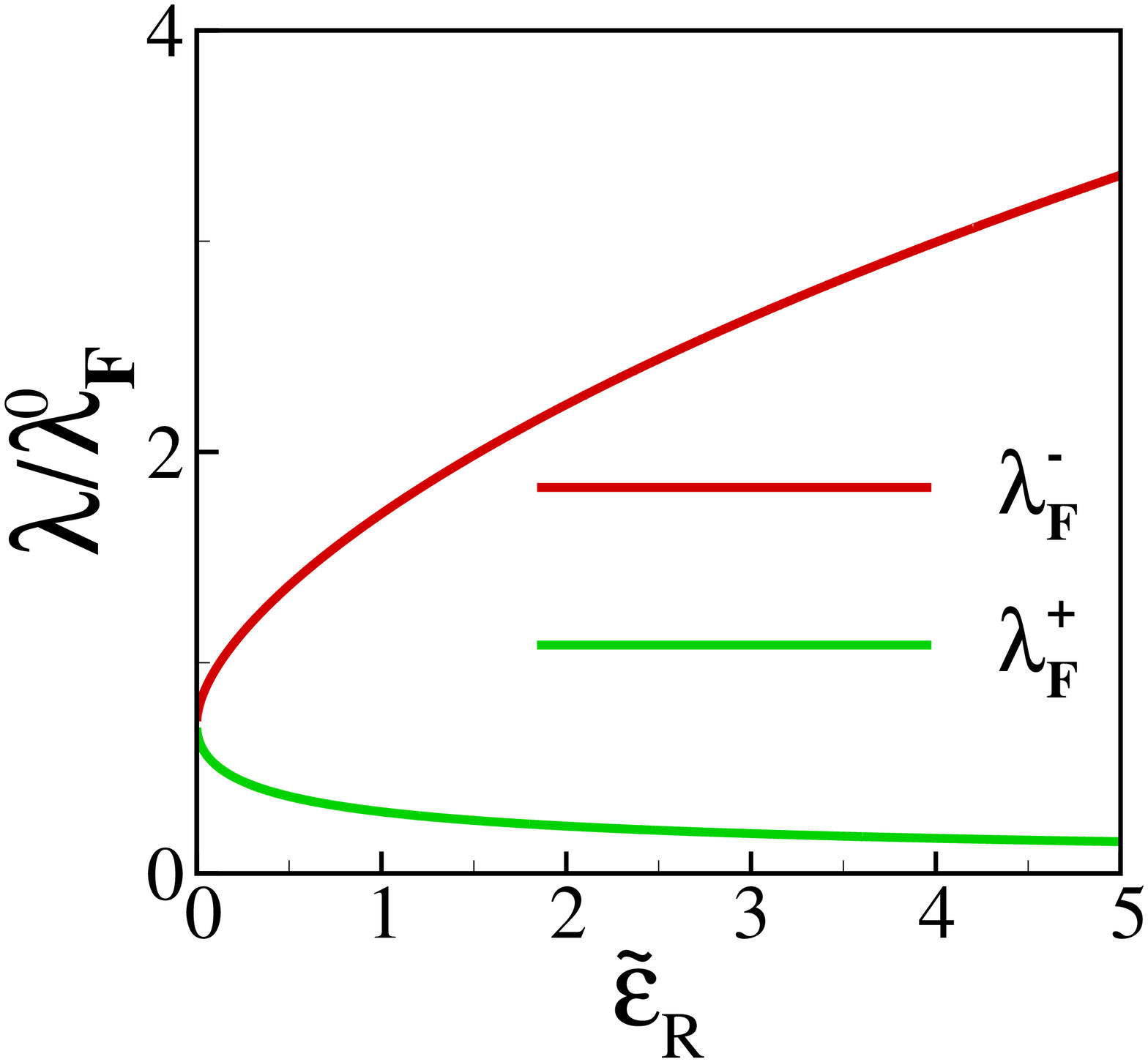}
   		\includegraphics[width=45mm]{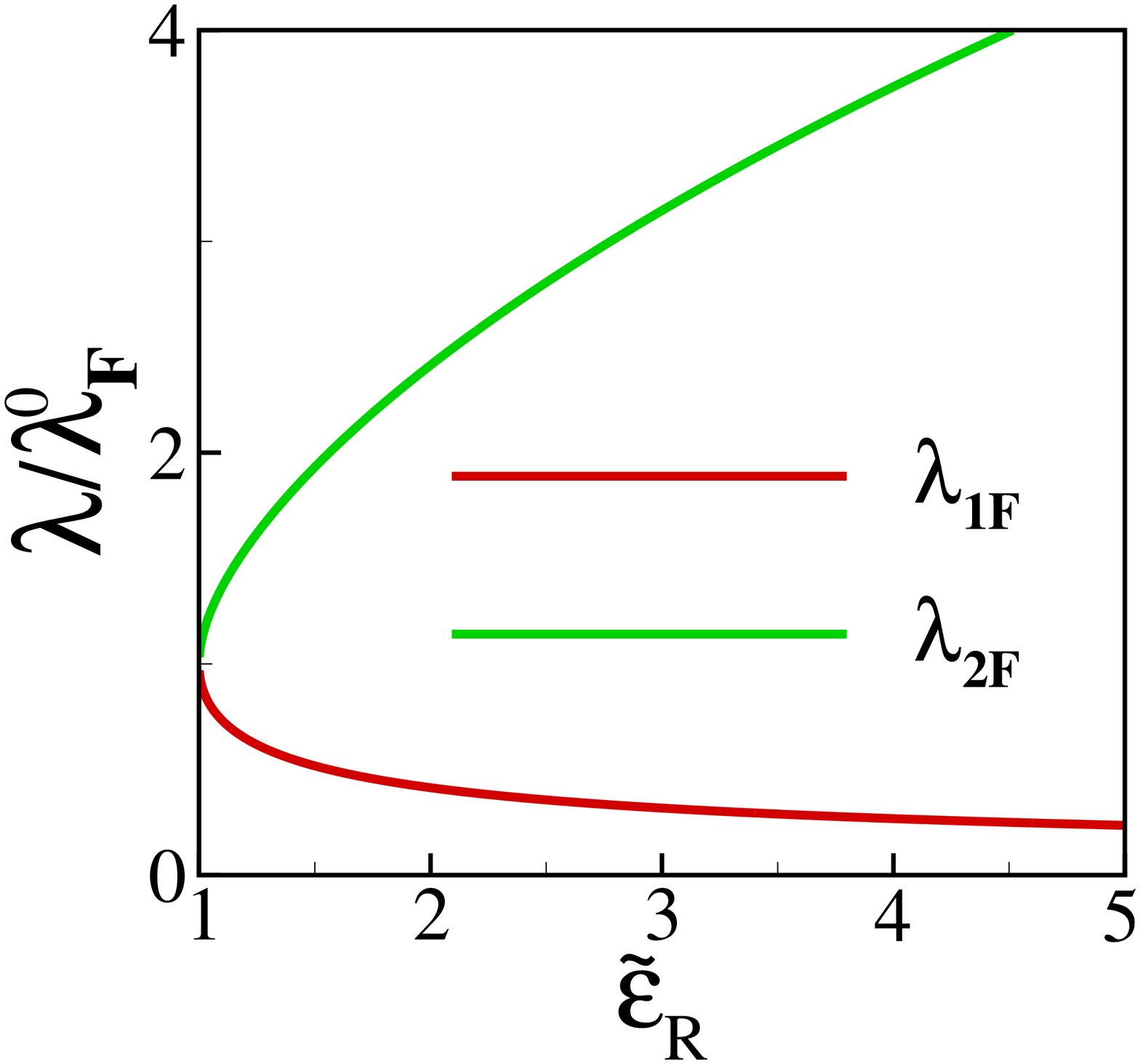}}
   	\caption{(Color online) Left: the Fermi wavelength of electrons in the bands $+$ and $-$, versus the strength of SOC, when Fermi energy is located above the BCP ($\ve_\f>0$). Right: the Fermi wavelength of electrons in the branches $1$ and $2$,  when Fermi energy is located below the BCP.}
   	\label{fig:Lambda}
   \end{figure}

When the spins of magnetic clusters are normal to the surface of the 2DRS ($\theta=0$), according to Eq. (\ref{two-band-w}) the scattering amplitude depends only on $\Delta \phi$ and the system is isotropic. In this case, the coefficients matrix in Eq. (\ref{mc}) reduces to a block diagonal matrix and the relaxation times $\boldsymbol\tau_x=(\tau_x^+,\tau_x^-)$ and $\boldsymbol\tau_y=(\tau_y^+,\tau_y^-)$ are simplified to 
\begin{equation}
\boldsymbol\tau_x=\boldsymbol\tau_y=\boldsymbol\tau={\bf A}_{0}^{-1}\cdot \d_1,
\end{equation}
where $\boldsymbol\tau=(\tau^+,\tau^-)$.
\begin{figure}[h]
\centerline{\includegraphics[width=70mm]{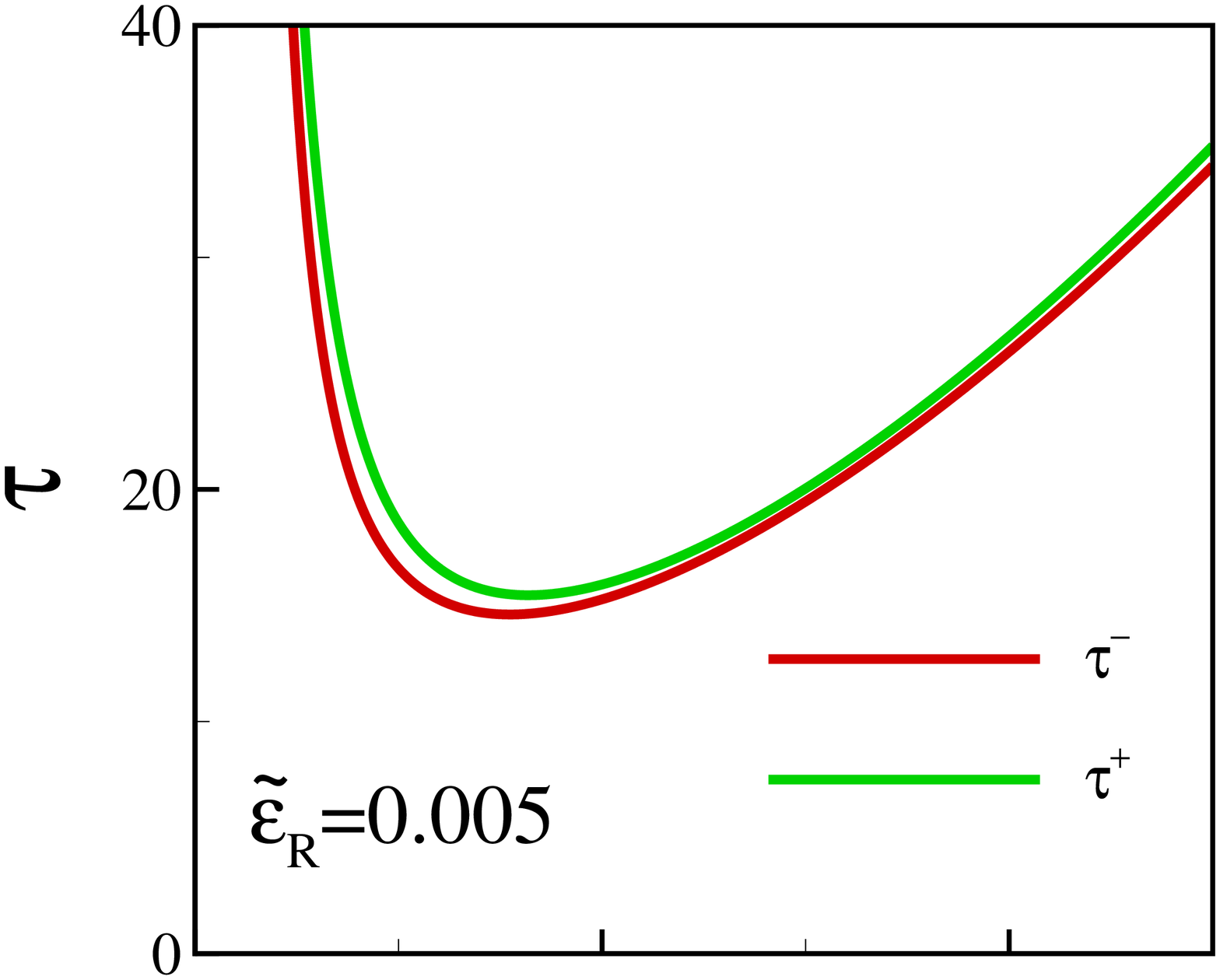}}
\vspace{-11mm}
\centerline{\includegraphics[width=70mm]{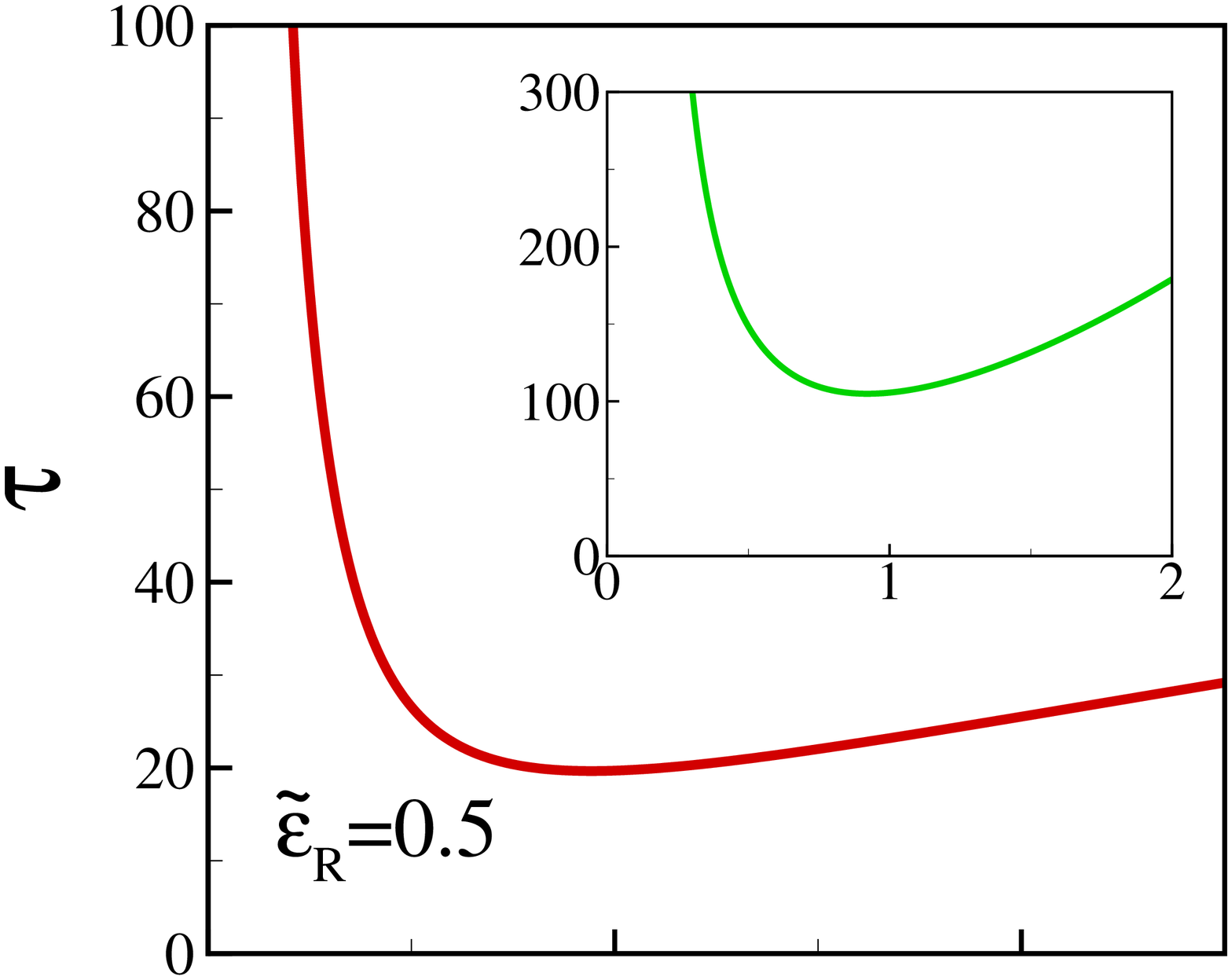}}
\vspace{-11mm}
\centerline{\includegraphics[width=70mm]{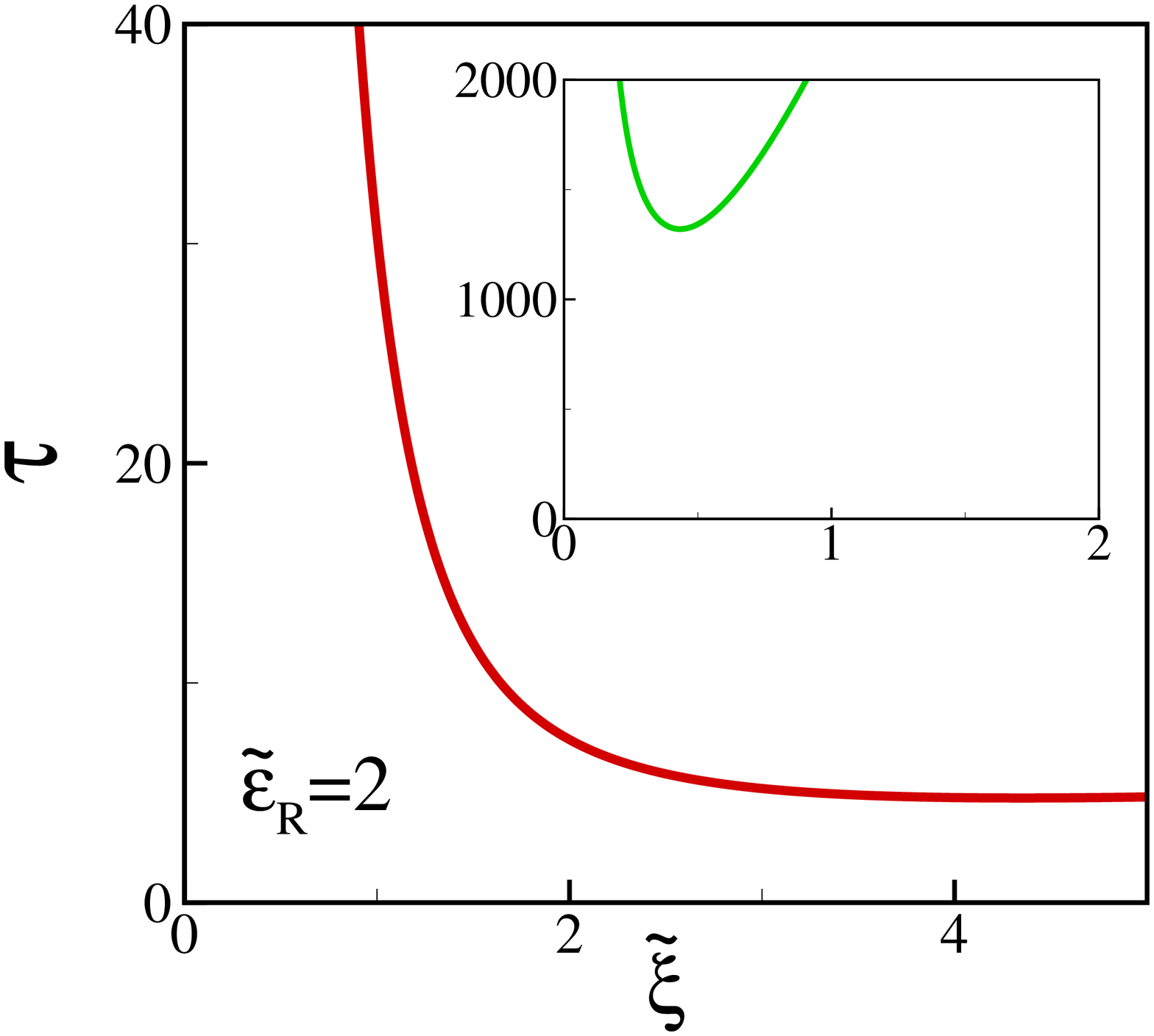}}
\caption{(Color online) The relaxation times of electrons in the bands $+$ and $-$ vs $\tilde{\xi}=\xi\sqrt{2m\ve_\f/\hbar^2}$, for different strengths of SOC, when the spins of magnetic clusters are normal to the surface of the 2DRS ($\theta=0$). Here, Fermi energy is located above the BCP.} \label{fig:time-two-band-isotrop}
\end{figure}
  We have plotted in Fig. \ref{fig:time-two-band-isotrop}, the relaxation times $\t^\pm$ versus CMS, for different strengths of SOC. 
  As it is seen both of them decrease by increasing $\tilde{\xi}$, become minimum and then gradually go to infinity at large CMSs.
   The emergence of such a minimum (which is 
   a combined effect of SOC and magnetic clusters) is
   attributed to the efficient scattering of electrons when their Fermi wavelength is comparable with CMS. 
  At a given Fermi energy, the Fermi wavelength of electrons in the band $n$ depends on the strength of SOC as $\lambda^n_\f=\la_\f^0/(\sqrt{\tver+1}+n \sqrt{\tver})$. 
 When CMS increases, maximum scattering of electrons in the band $n$ occurs at $\la_\f^n\sim 2\pi\xi$.   
  Since the Fermi wavelengths in the bands $+$ and $-$ are not the same size (see Fig. \ref{fig:Lambda}), minimums of $\tau^\pm$ appear at different CMSs. 
  For small SOCs the Fermi wavelength $\la_\f^n$ behaves as $\la_\f^n\sim\la_\f^0(1-n\sqrt{\tver})$, and the separation between minimums increases by SOC as $\sqrt{\tver}$. For large SOCs the minimum of $\t^+$ approaches gradually to the point $\tilde{\xi}=0$, but $\t^-$ becomes minimum at larger CMSs.
  Also, since $\la^+_\f$ is smaller than $\la^-_\f$, the relaxation time $\tau^+$ is always larger than $\t^-$.
  
\begin{figure}[h]
          \centerline{\includegraphics[width=70mm]{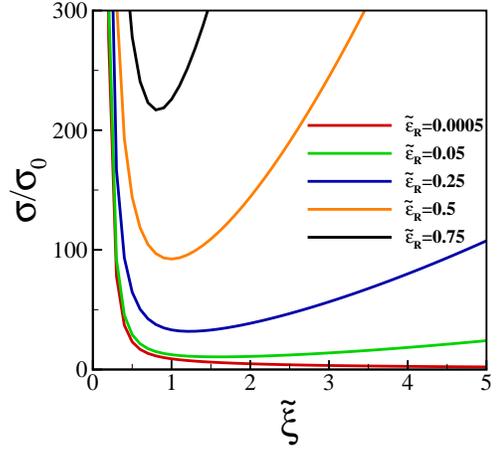}}
          \caption{(Color online) The conductivity of the isotropic 2DRS ($\theta=0$) vs CMS for different strengths of SOC. At a given Fermi energy above the BCP ($\ve_\f>0$), by increasing the strength of SOC, the conductivity increases and it's minimum emerges at smaller values of $\tilde{\xi}$.} \label{fig:conductivity-isotrop-x-2-band}
\end{figure}

In the isotropic case since the relaxation times $\tau^{n}_x$ and $\tau^{n}_y$ are equal, so are the conductivities along $x$ and $y$ directions ($\sigma_{xx}=\sigma_{yy}=\sigma$).
For small CMSs the conductivity is very large, however by increasing CMS it decreases rapidly, becomes minimum at a $\tilde{\xi}$ (let say $\tilde{\xi}_{min}$), and finally increases monotonically, as shown in Fig. \ref{fig:conductivity-isotrop-x-2-band}.
To describe this non-monotonic behavior, we investigate the behavior of the scattering amplitude. Let us write the $T$-matrix as 
 \begin{equation}
 T^{n,n'}_{\kv,\kv'}=T^{\theta}_{n,n'}T^{\xi}_{\kv,\kv'},\label{t-parts}
 \end{equation}
which is a multiple of two parts: 1) the spin-dependent part: $T^{\theta}_{n,n'}$, which depends on the tilt angle $\theta$. This part is given by:
\begin{equation}
T^{\theta}_{n,n'}=\cos\theta(n n'e^{i\Delta\phi}-1)-\sin\theta(n'e^{-i\phi'}+n e^{i\phi}).
\label{teta-depend}
\end{equation}
2) The $\xi$-dependent part, written as
\begin{equation}
T^{\xi}_{\kv,\kv'}=\int d\rv e^{-i\kv\cdot\rv}e^{-r/\xi}e^{i\kv'\cdot\rv},
\label{kisi-depend}
\end{equation}
which depends on the clusters' mean size, $\xi$. 

For large CMSs, in the limit of $\xi\rightarrow\infty$, the $\xi$-dependent part of the $T$-matrix becomes proportional to the Dirac delta function $\delta(\kv-\kv')$, and consequently only intra-bands forward scatterings have determinant effects on the $T$-matrix. 
On the other hand, from the spin dependent part we see that at $\theta=0$, the intra-band scattering amplitudes, $T_{+,+}^{\theta}$ and $T_{-,-}^{\theta}$, are equal to $(\exp(i\Delta\phi)-1)$. These amplitudes are vanishing for $\Delta\phi=0$, which means that no intra-band forward scattering occurs in the system. 
Summing up the above arguments, we conclude that at $\theta=0$, for $\xi\rightarrow\infty$ the scattering amplitude approaches to zero which results in an infinite conductivity.

For small CMSs, in the limit of $\xi \rightarrow 0$, the scattering potential becomes weaker, and thus no scattering happens in the system and the conductivity is very large. 

We can also explain the physics behind the non-monotonic behavior of the conductivity by comparing the average Fermi wavelength of electrons obtained as 
\begin{eqnarray}
\nonumber\la^{av}_\f&=&\frac{N_+}{N_++N_-}\la_\f^++\frac{N_-}{N_++N_-}\la_\f^-\\
&=&\frac{\la_\f^0}{\sqrt{1+\tilde{\ve}_\r}}\label{Eq:lambda-Av},
	\end{eqnarray}
with the CMS, $\xi$. 
At a given $\tilde{\ve}_\r$, when CMS increases, backscattering of electrons by clusters increases until $\xi$ becomes comparable with $\la_\f^{av}$. When $\xi\sim\la_\f^{av}/2\pi$ the ability of clusters to scatter electrons is greatly increased and electrons are most efficiently scattered by magnetic clusters. At this point the conductivity becomes minimum.
But for CMSs larger than $\la_\f^{av}$, dominant scatterings by clusters' size are forward, however spin of clusters try to scatter electrons in a backward direction. When CMS increases these two effects cancel each other, and the conductivity enhances monotonically by increasing CMS.

Moreover, the location of the minimum of the conductivity depends on the strength of SOC. In the absence of SOC, the Rashba system (\ref{eq:hamiltonian}) reduces to a free electron system and the conductivity always decreases by increasing $\tilde{\xi}$. However, when SOC increases, due to the relation in Eq. (\ref{Eq:lambda-Av}) the average Fermi wavelength becomes smaller and thus $\tilde{\xi}_{min}$ decreases. 

\begin{figure}[h]
             \centerline{\includegraphics[width=50mm]{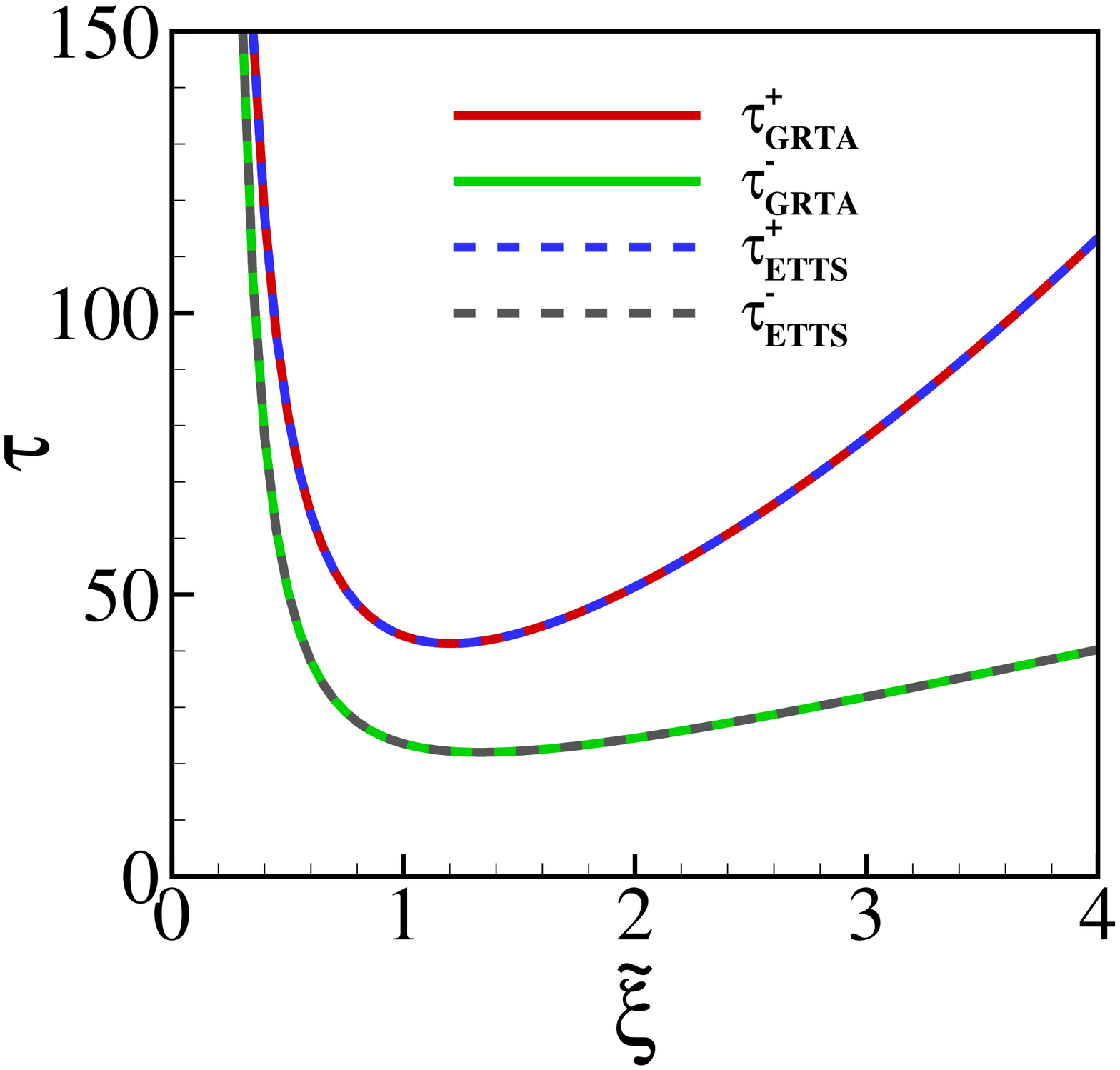}
             \hspace{-5.0mm}
             	\includegraphics[width=50mm]{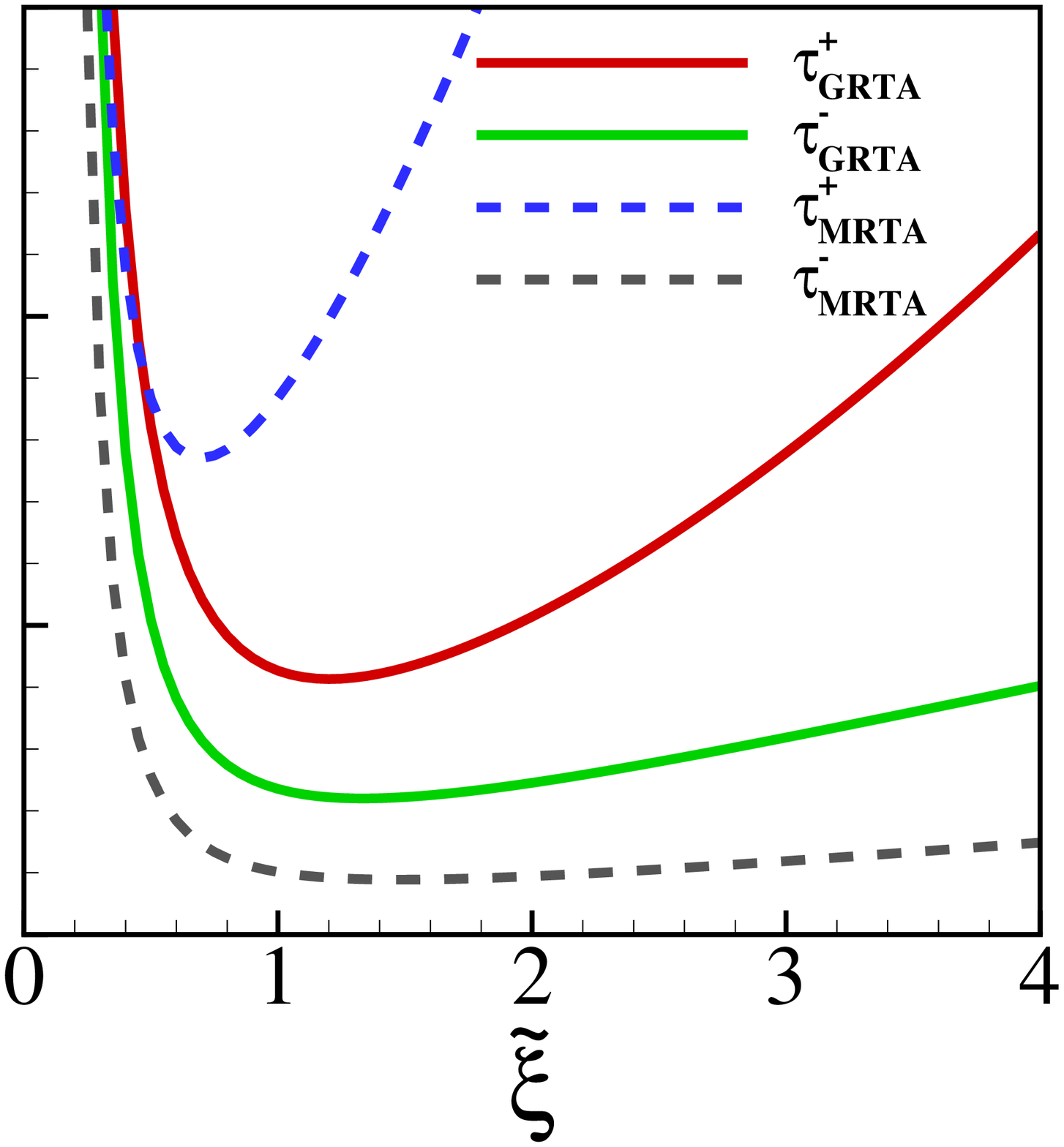}}
             \caption{(Color online) The relaxation times in the bands $+$ and $-$ vs CMS when $\tver=0.25$. Left: the dashed lines are the results of ETTS method, and the solid lines are obtained by GRTA. Right: the dashed lines are the results of MRTA, and the solid lines are obtained by GRTA.} \label{fig:ETTS-MRTA-GRTA}
             \end{figure}

 {\it A comparison with other approaches}:  
 
 So far, several methods have been proposed to solve the Boltzmann equation for isotropic 2DRSs. In the method presented by C. Xiao, et. al, by developing the Boltzmann technique they established a set of self-consistent equations for the transport times \cite{Xiao2016,PhysRevB.93.075150}. Their calculations are based on the exact transport time solution (ETTS) of the Boltzmann equation in the Born approximation. In this method, by introducing an isotropic transport time for electrons with energy $\ve$ in the band $n$, via
 \begin{equation}
 f_{n}-f_{n}^0=\left(\frac{\partial f^{0}_n}{\partial\ve_n}\right){\bf E}\cdot{\bf v}_{n}(\phi)\tau^{n}(\ve),
 \end{equation}
and using Boltzmann equation, we obtain the following self-consistent equation for the relaxation times 
 \begin{equation}
 \begin{aligned}
 &\frac{1}{\tau^{n}_{ETTS}(\ve_\f)}=\\
 &\sum_{n'=\pm}
 \int\frac{d\phi'}{2\pi}w_{n,n'}(\varepsilon_\f)
 [1-\cos(\Delta\phi)\frac{v_{n'}(\phi')}{v_{n}(\phi)}\frac{\tau^{n'}(\varepsilon_\f)}{\tau^{n}(\varepsilon_\f)}].
 \end{aligned}
 \end{equation}
 By solving this integral, we reach to two linear equations for the relaxation times $\tau^{+}$ and $\tau^{-}$ (not shown). In Fig. \ref{fig:ETTS-MRTA-GRTA} (left panel), we have plotted the relaxation times obtained by ETTS method. The results of GRTA are also plotted for comparison. As it is seen, these two methods are completely consistent. 
             
 Modified relaxation time approximation (MRTA) is another method for solving the Boltzmann equation for isotropic 2DRSs. In this method, the relaxation time in the band $n$ is given by: 
 \begin{equation}
 \begin{aligned}
&\frac{1} {\tau^{n}_{MRTA}(\varepsilon_\f)}=\sum_{n'=\pm}
\int\frac{d\phi'}{2\pi}w_{n,n'}(\varepsilon_\f)
[1-\cos(\Delta\phi)\frac{v_{n'}(\phi')}{v_{n}(\phi)}].
\end{aligned}
 \end{equation}    
 In both the ETTS and GRTA methods, we need to solve two coupled equations in order to obtain the relaxation times, while in the MRTA method band dependence of $\t_{MRTA}$ is neglected.
 In Fig. \ref{fig:ETTS-MRTA-GRTA} (right panel), we have plotted the relaxation times obtained by the MRTA. In the band $-$, due to the larger Fermi wavelength, we expect the relaxation time to be smaller, but in GRTA, because of the coupling between the two bands, the relaxation time $\t^-$ is larger than the corresponding one in MRTA. Similarly, in the band $+$, because of the coupling with the band $-$, the relaxation time is smaller than the corresponding one in MRTA.
  Regardless of the differences in these methods, the behaviors of the relaxation times versus CMS are qualitatively the same. The same discussions are also valid when the Fermi energy is located below the BCP.       

\subsection{Anisotropic magnetoresistance (AMR)}

\begin{figure}[h]
                \centerline{\includegraphics[width=45mm]{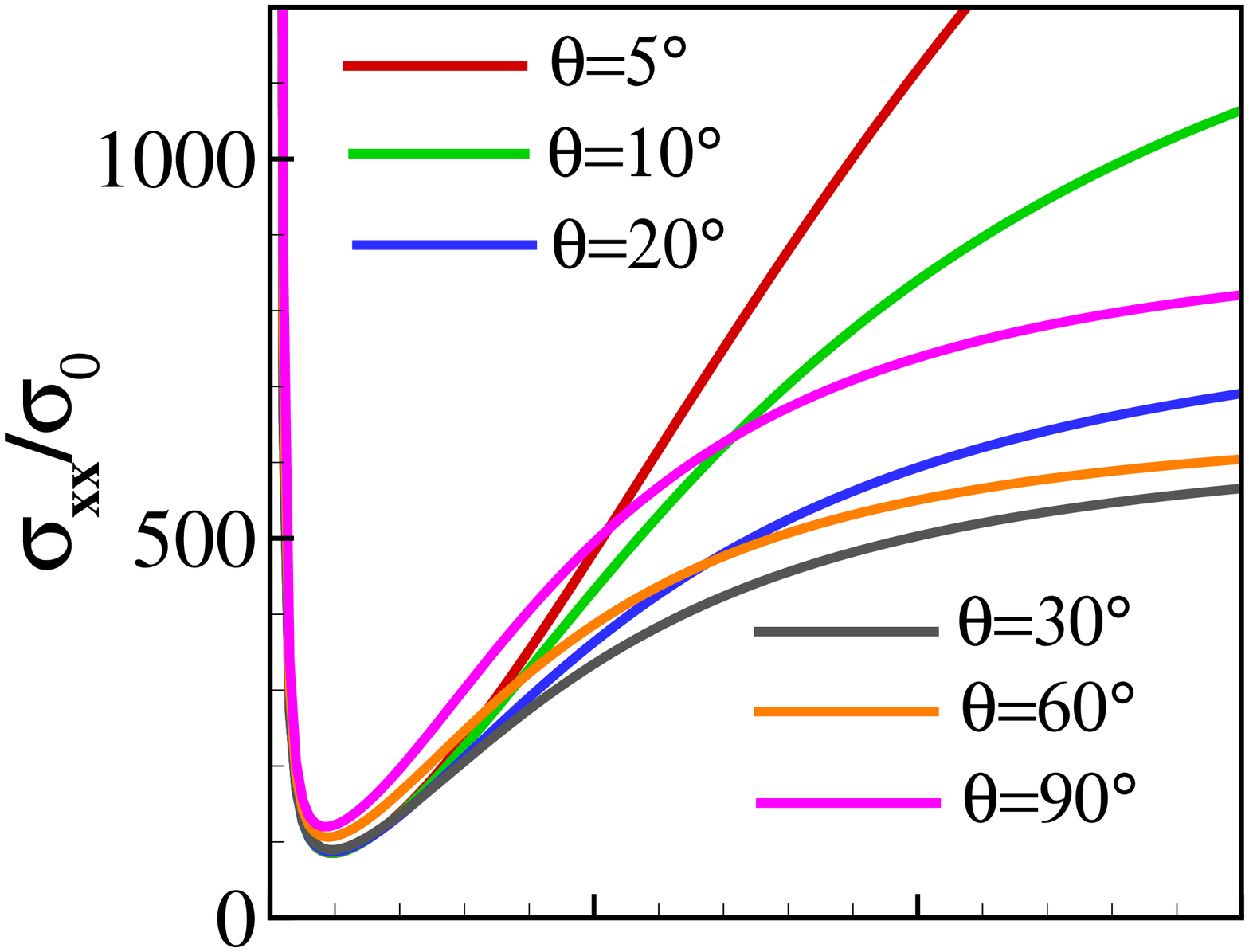}\vspace{-0.25mm}\includegraphics[width=45mm]{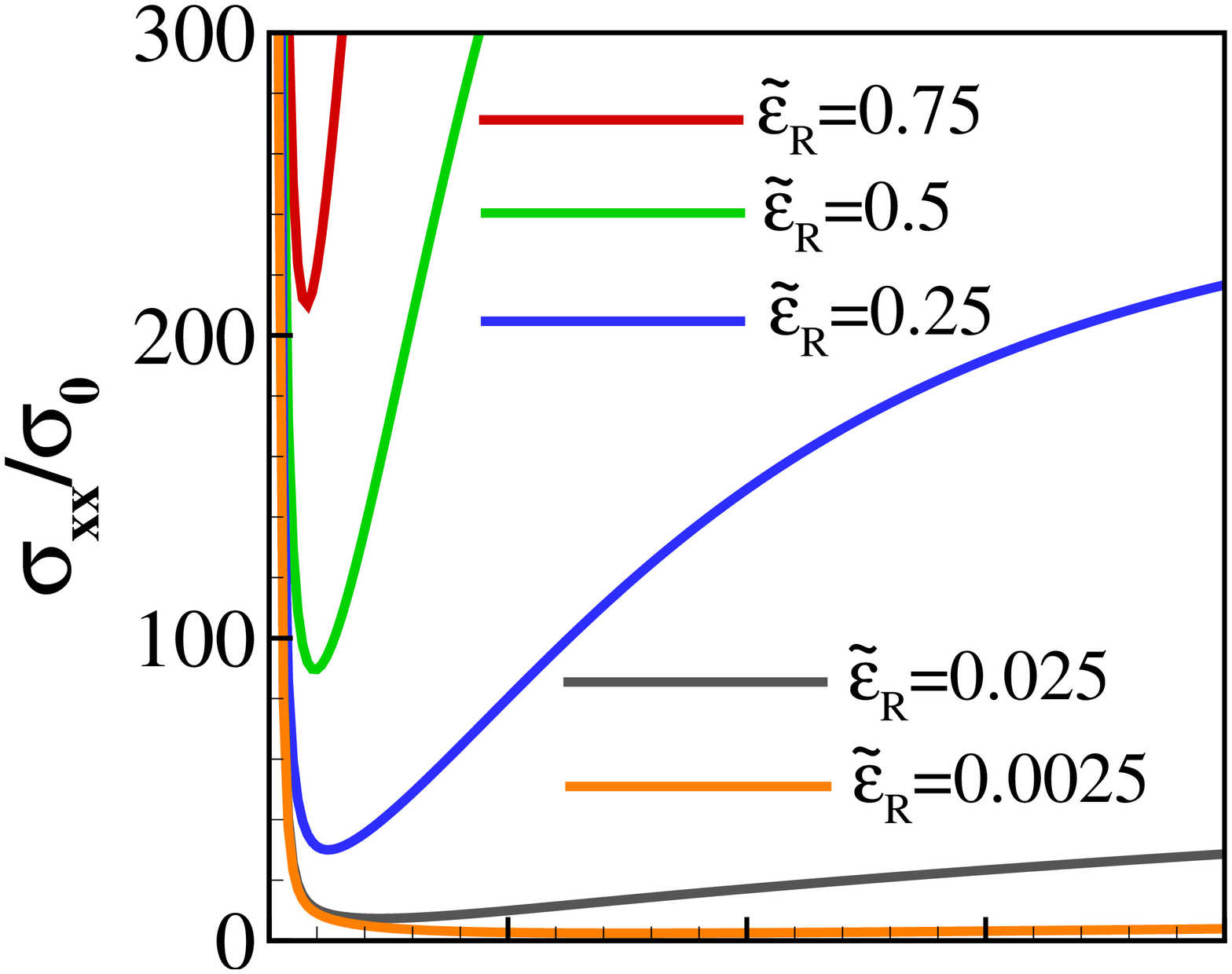}}
                \centerline{\includegraphics[width=45mm]{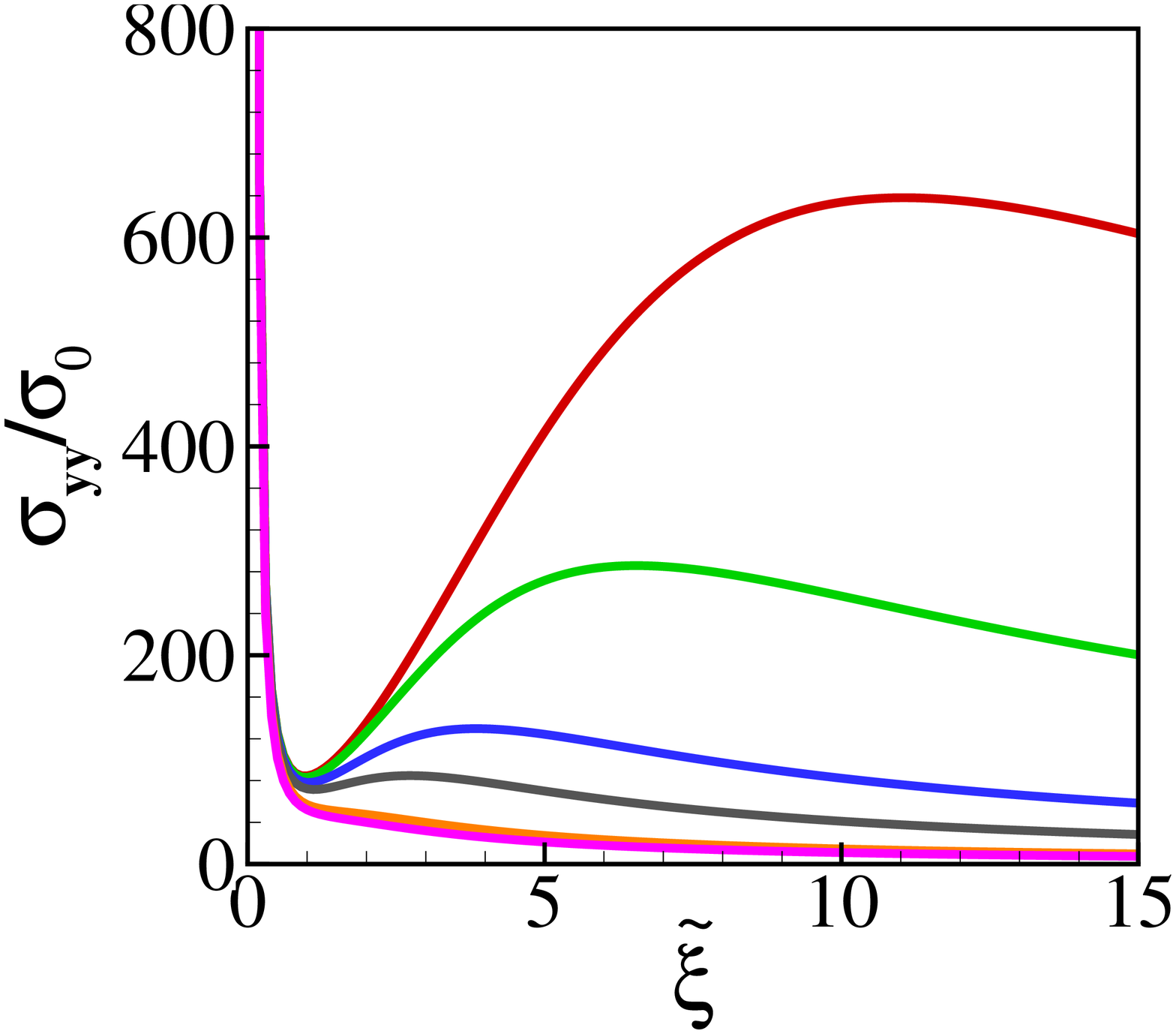}\vspace{-0.25mm}\includegraphics[width=45mm]{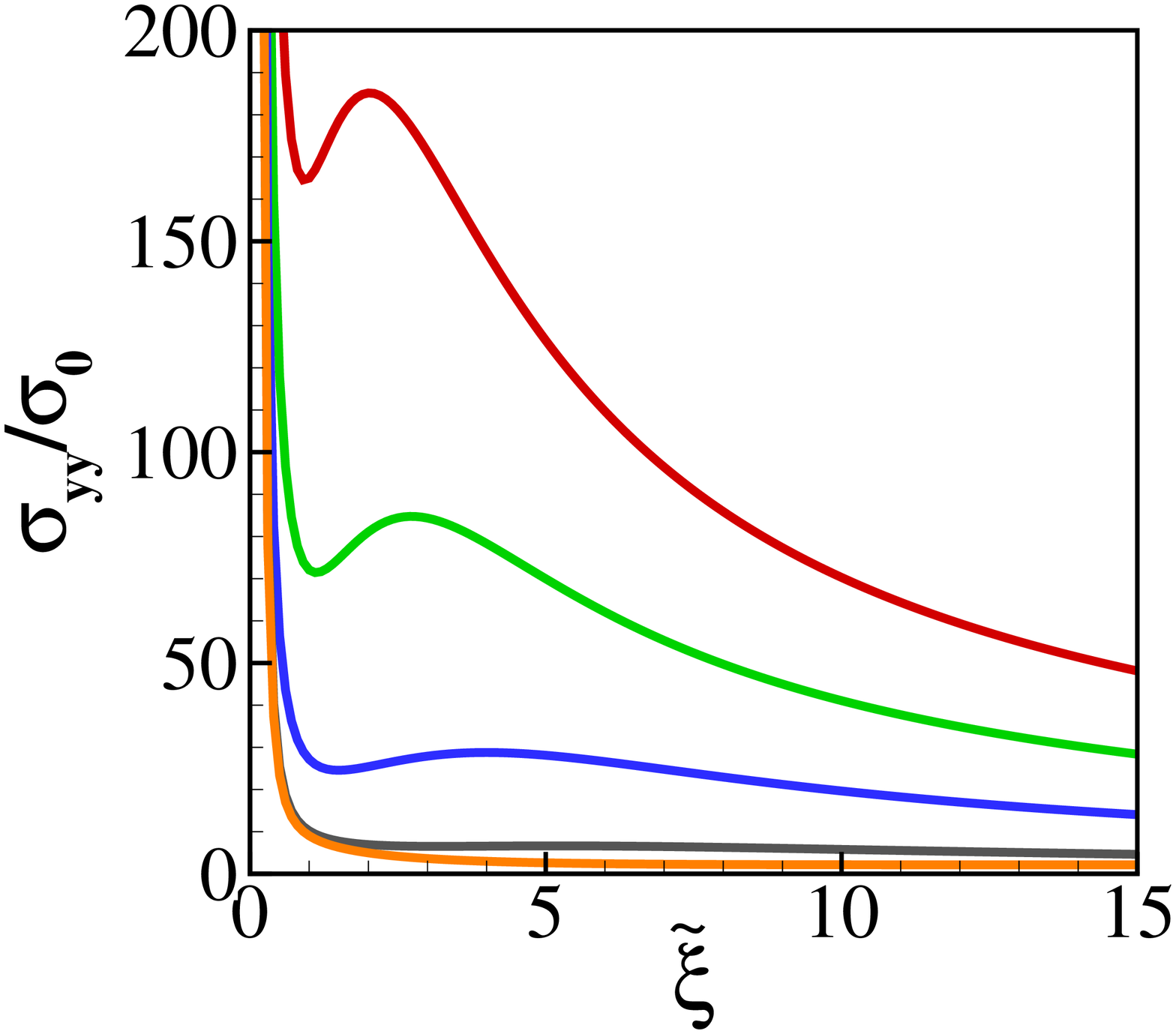}}
                \caption{(Color online) Left column: the conductivity of the 2DRS along $x$ and $y$ directions vs CMS, for different values of the tilt angle $\theta$, when $\tver=0.5$. Right column: the conductivity of the 2DRS along $x$ and $y$ directions vs CMS, for different strengths of SOC, when $\theta=\pi/6$. Here, Fermi energy is located above the BCP.} \label{fig:sigmaxx-yy-2-band}
         \end{figure}

   When, the spins of magnetic clusters are aligned in a direction given by $\theta$ (see Fig. \ref{fig:schematic}), the system is highly anisotropic and the conductivities in $x$ and $y$ directions behave, differently. Unlike the isotropic case, the conductivity $\sigma_{xx}$ increases after the minimum, and saturates at large CMSs (see Fig. \ref{fig:sigmaxx-yy-2-band}), however $\sigma_{yy}$ has an oscillatory-like behavior, it decreases by increasing CMS as shown in Fig. \ref{fig:sigmaxx-yy-2-band}. 
   Such a different behavior of $\sigma_{xx}$ and $\sigma_{yy}$ is attributed to the exchange interaction of electrons with magnetic clusters.
   When the electric field is applied along $x$ axis, most of electrons involved in $\sigma_{xx}$ move in the $x$ direction, and due to the spin-orbit locking their spins lie in $y$ direction. By increasing the tilt angle $\theta$, $S_y$ becomes larger and due to the electron-cluster exchange interaction the spin accumulation along $y$ axis increases which, in turn, causes the increase of conduction electrons along $x$ direction.
      But, for conductivity along $y$ axis, electrons' spins are perpendicular to $S_y$. When $S_y$ becomes larger, electrons try to align their spin with $S_y$ to minimize the exchange energy. Again due to the SOC, electrons have to change their direction of motion which leads to a reduction of $\sigma_{yy}$.
        
\begin{figure*}[t]
  \centerline{\includegraphics[width=50mm]{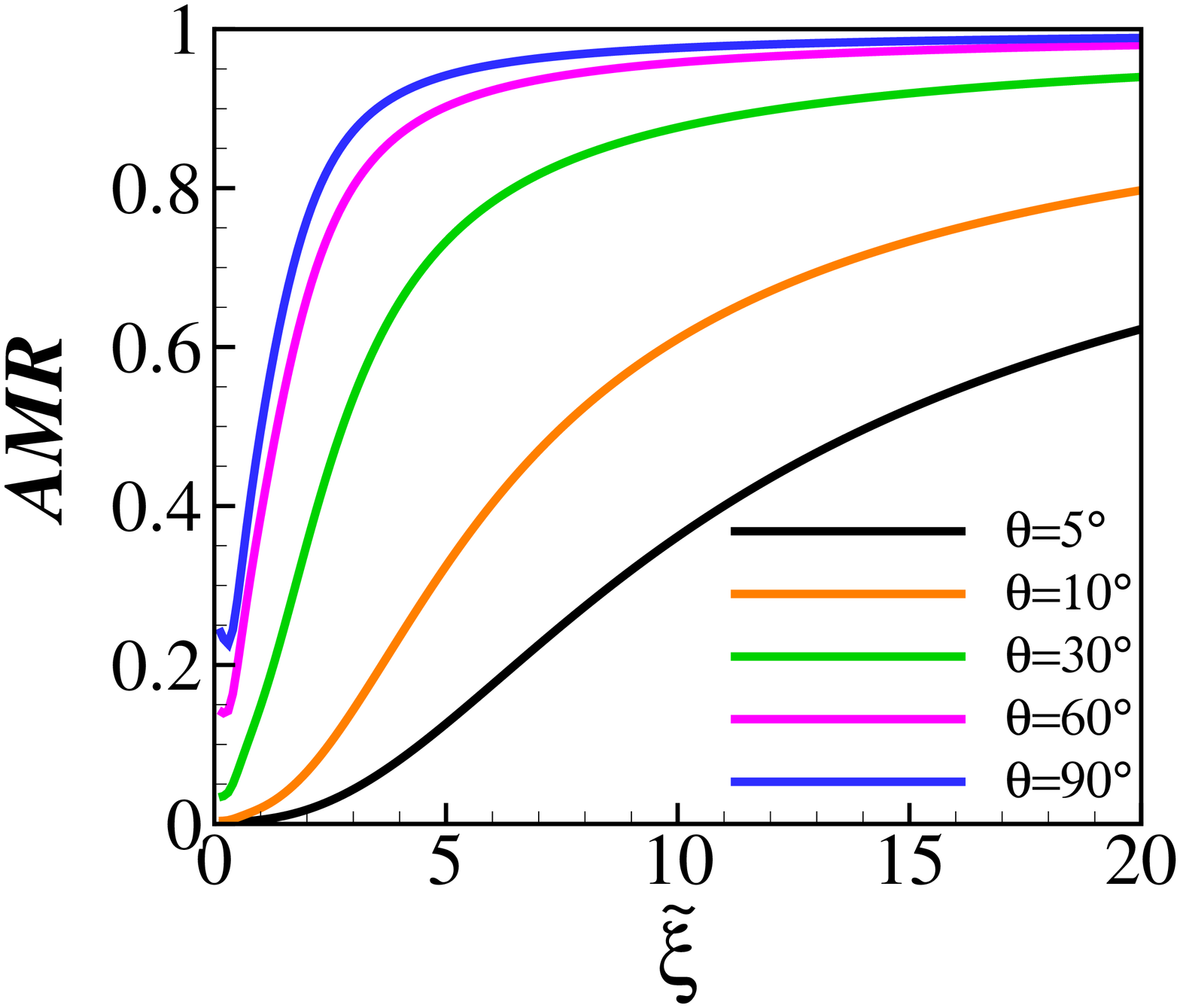}\hspace{-0.3mm}
    \includegraphics[width=50mm]{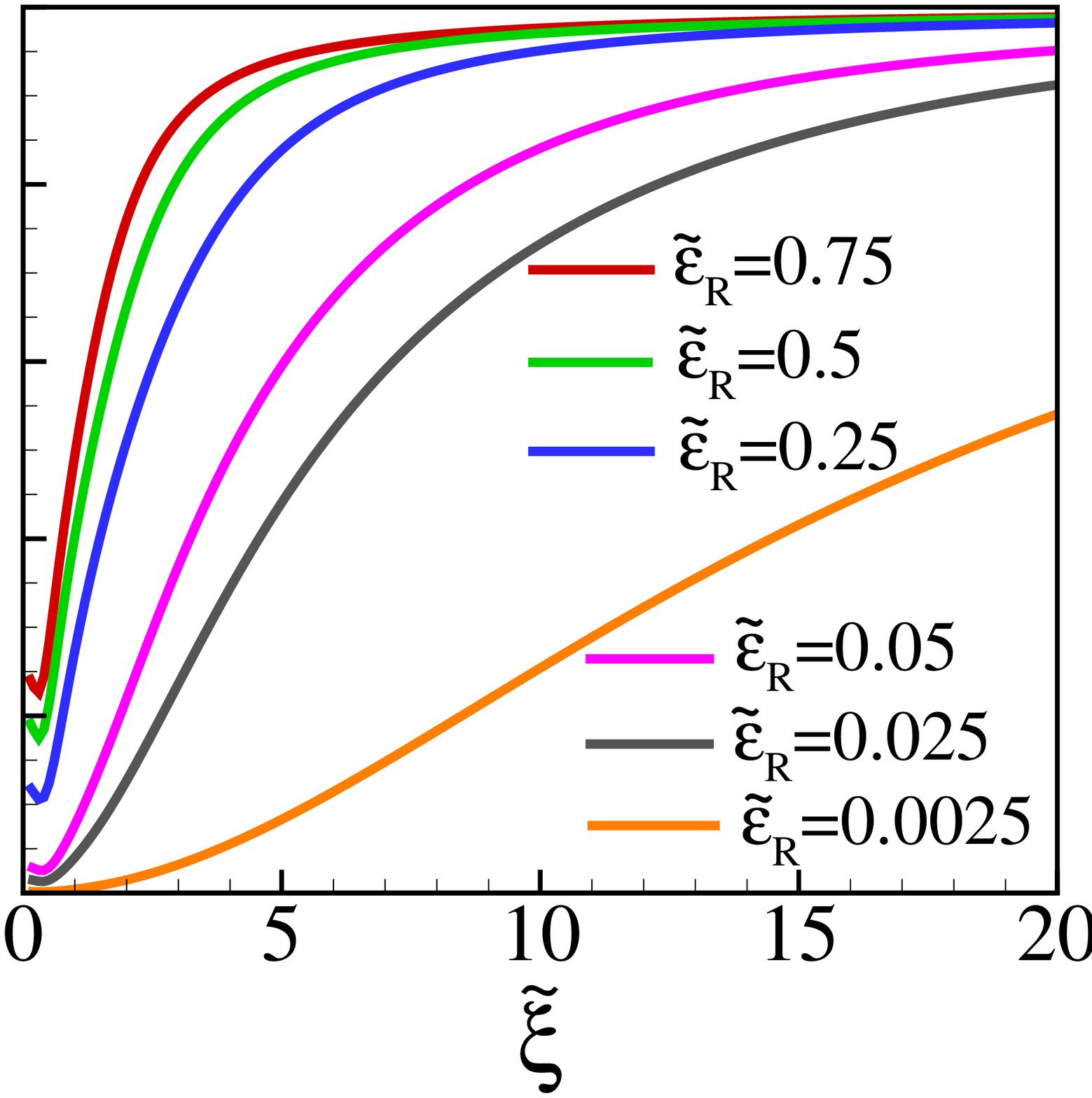}\hspace{-8.3mm}
\includegraphics[width=50mm]{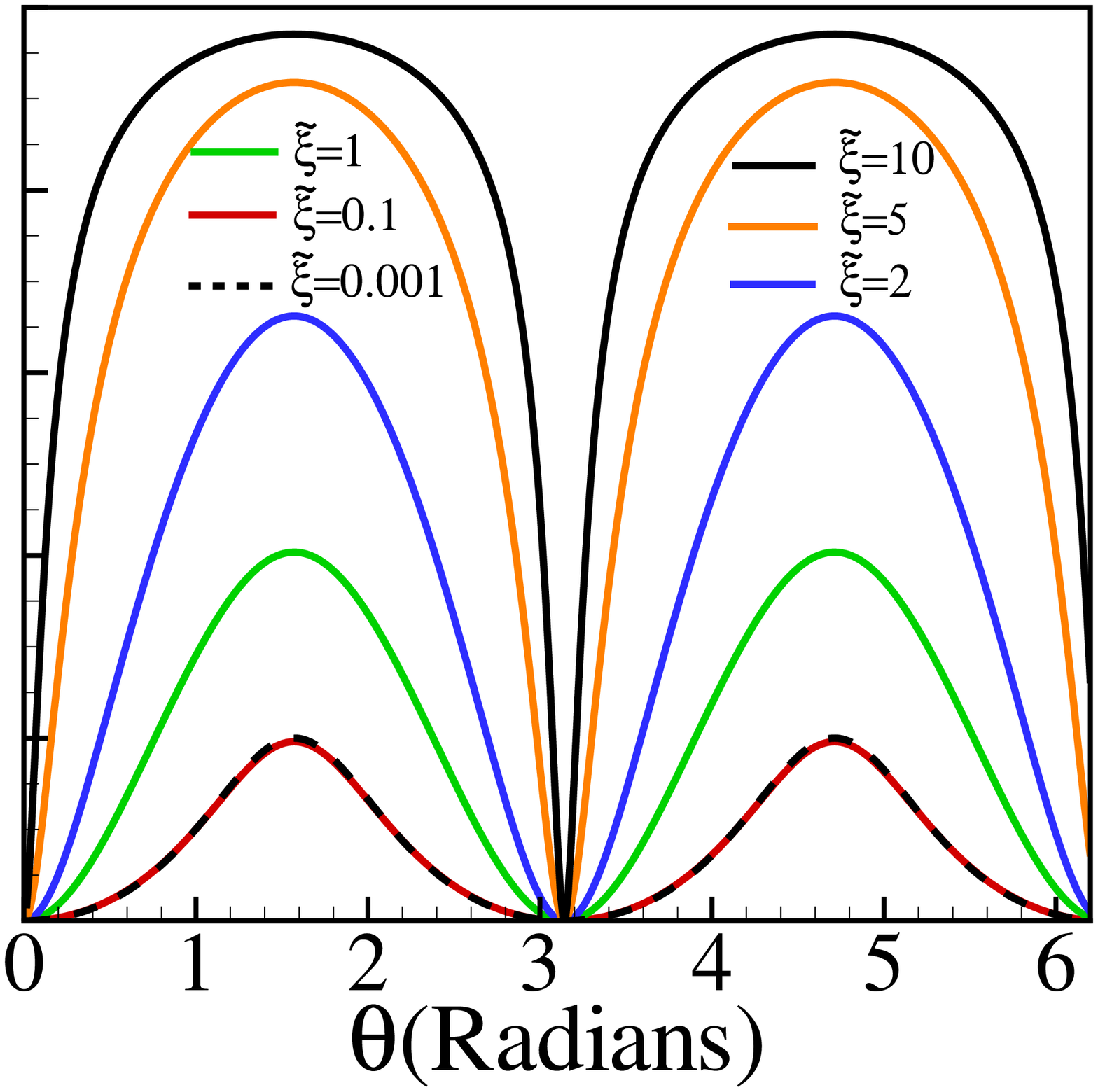}\hspace{-8.3mm}
\includegraphics[width=50mm]{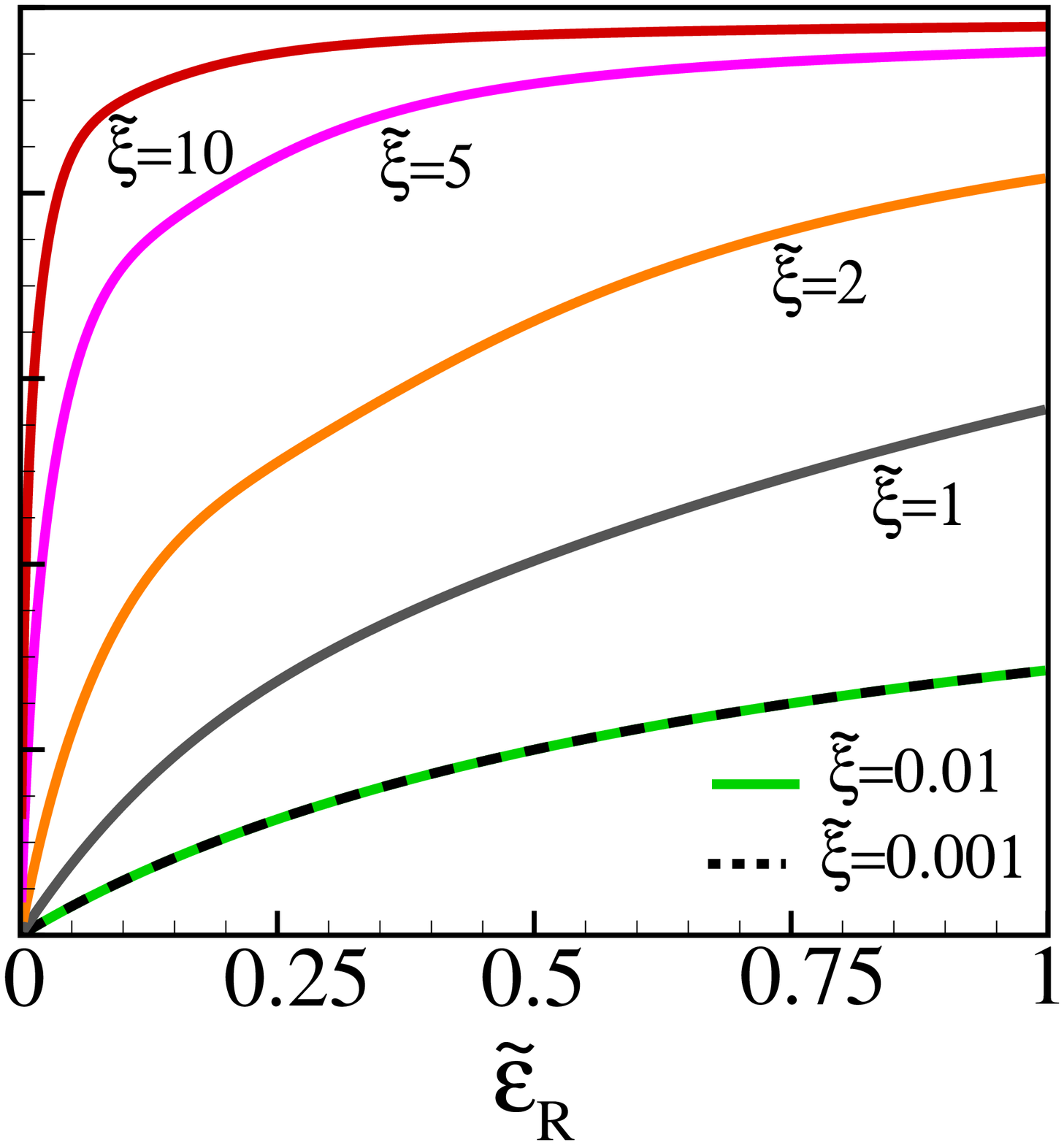}}
 \caption{(Color online) From left to right: The AMR vs CMS for different values of the tilt angle $\theta$, when $\tver=0.75$. The AMR vs CMS for different strengths of SOC, when $\theta=\pi/2$. The AMR vs the tilt angle $\theta$ for different CMSs, when $\tver=0.5$. The AMR vs $\tver$ for different CMSs, when $\theta=\pi/2$.}
  \label{fig:AMR-2-band}
  \end{figure*}

As an amount of the anisotropy of the system, we investigate the behavior of AMR, defined as: 
    \begin{equation}
    AMR=\frac{\sigma_{xx}-\sigma_{yy}}{\sigma_{xx}+\sigma_{yy}}.
    \end{equation}
In Fig. \ref{fig:AMR-2-band}, we have plotted the AMR versus CMS, for different values of the tilt angle $\theta$, when the strength of SOC is $\tver=0.75$. By increasing $\tilde{\xi}$, the AMR increases monotonically and saturates to unit at large values of $\tilde{\xi}$. Also, by increasing the tilt angle $\theta$ the AMR increases, implying that the system is more anisotropic when clusters' spin lies on the surface of the 2DRS.
We have also plotted in Fig. \ref{fig:AMR-2-band}, the AMR versus $\tilde{\xi}$ for different strengths of SOC when $\theta=\frac{\pi}{2}$. As it is seen, by increasing $\tver$ the anisotropy of the system enhances (this behavior, i.e. the increase of AMR by increasing the strength of SOC, occurs only for the regime of $\ve_\f>0$. In the regime of $\ve_\f<0$ the AMR experiences a minimum at a $\tver$, which will be discussed in next section).       
In order to see the angular dependence of the AMR, we have also plotted in Fig. \ref{fig:AMR-2-band}, the AMR versus $\theta$, for different values of CMS, at $\tver=0.5$. As it is seen the angular dependence of the AMR changes by varying $\tilde{\xi}$. The unconventional behavior of the AMR as a function of the tilt angle $\theta$ is a combined effect of SOC and magnetic clusters.  
The maximum value of the AMR increases by increasing $\tver$. To show this dependence for very small CMSs, we have also plotted in Fig. \ref{fig:AMR-2-band}, the AMR versus $\tver$ for different CMSs, at $\theta=\pi/2$. As it is clearly seen the anisotropy exists even at very small SOCs, it grows monotonically by increasing SOC and saturates to 1. This property is not seen in the regime of $\ve_\f<0$. As we will show in next section, below the BCP the AMR is almost constant with respect to $\tver$ for small CMSs.
  
The behavior of the conductivity in 2DRSs is generally similar to the surface conductivity of 3D magnetic topological insulators.\cite{PhysRevB.98.155413} However, in contrast to the topological insulators, in 2DRSs $\tilde{\xi}_{min}$ strongly depends on the strength of SOC (see Fig. \ref{fig:sigmaxx-yy-2-band}). By decreasing the SOC, $\tilde{\xi}_{min}$ increases and goes to infinity in the limit of $\tilde{\xi}\rightarrow 0$. 
 
 \section{Single band scattering ($\ve_\f<0$)} \label{sec:single-band}
 
\begin{figure}[h]
                              \centerline{\includegraphics[width=45mm]{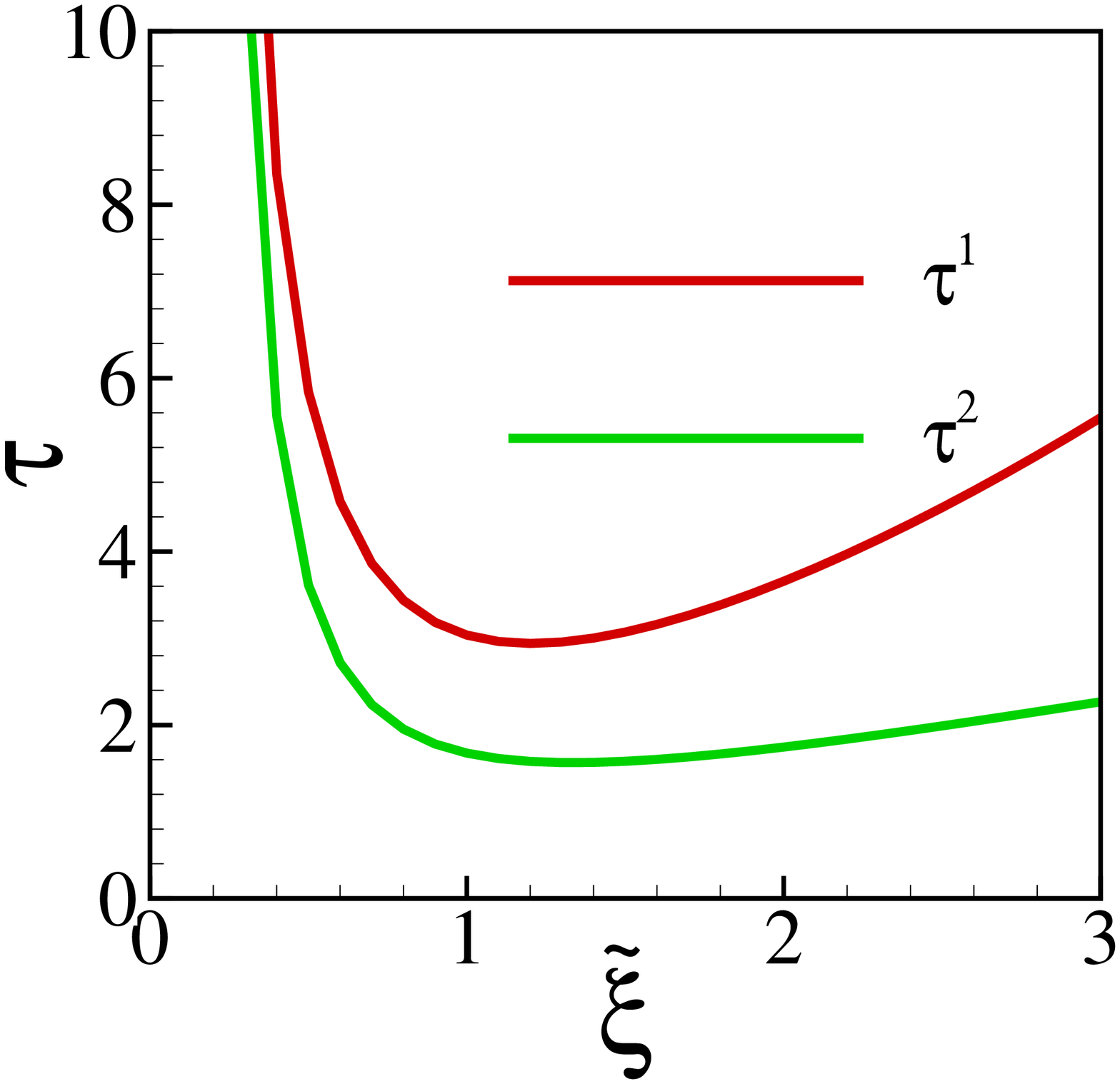}\includegraphics[width=45mm]{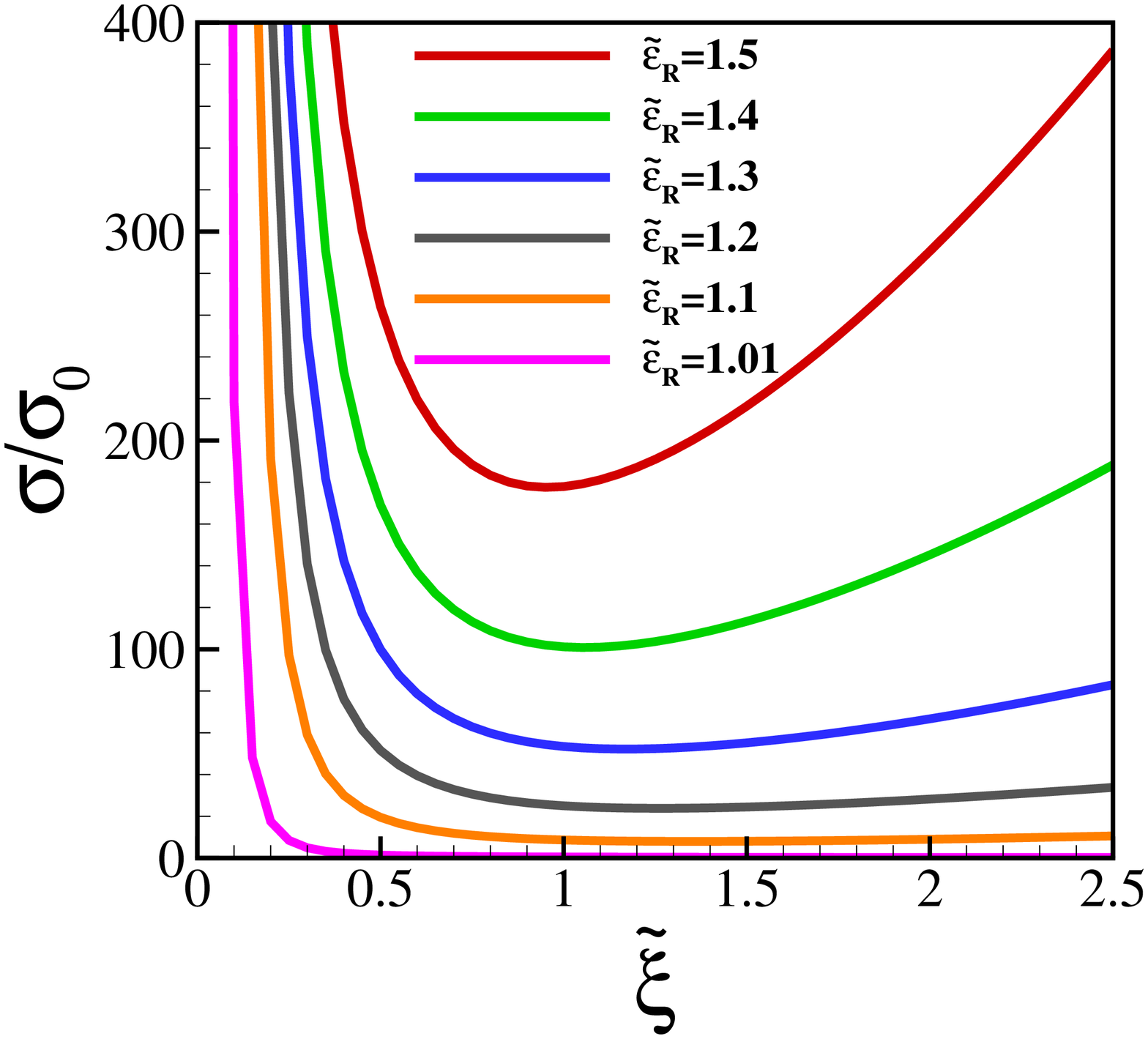}}
                              \caption{(Color online) Left: the relaxation times of electrons in the branches 1 and 2 vs CMS, when $\theta=0$, and $\tver=1.25$. Right: the conductivity of the isotropic 2DRS for different strengths of SOC, when Fermi energy is located below the BCP ($-\ve_\r/2<\ve_\f<0 $). } \label{fig:conductivity-isotrop-x-single}
                               \end{figure}

Below the BCP, the conductivity of the 2DRS is obtained as (see appendix \ref{Appendix-single-band}):
\begin{equation}
       \begin{aligned}
       &\sigma          _{xx}=\sigma_{0}\sqrt{\tver-1}\sum_{\nu=1,2}\tau_{x}^\nu(|\ve_\f|)\Big[\sqrt{\tver}-(-1)^{\nu}\sqrt{\tver-1}\Big],\\
       &\sigma          _{yy}=\sigma_{0}\sqrt{\tver-1}\sum_{\nu=1,2}\tau_{y}^\nu(|\ve_\f|)\Big[\sqrt{\tver}-(-1)^{\nu}\sqrt{\tver-1}\Big],
        \label{sigma-single-band-iso}
       \end{aligned}
       \end{equation}
where $\tver=\ve_\r/(2|\ve_\f|)$.   
In the isotropic case, $\t_x^1=\t_y^1=\t^1$ and $\t_x^2=\t_y^2=\t^2$, and the conductivities along $x$ and $y$ directions are identically the same. 
In Fig. \ref{fig:conductivity-isotrop-x-single} (left panel), we have plotted the relaxation times $\t^{1,2}$ versus $\tilde{\xi}$. 
Since the Fermi wavelength in the branch 1 is smaller than 2 (see Fig. \ref{fig:Lambda}), the relaxation time $\t^1$ is always larger than $\t^2$. The Fermi wavelength in the branch $\nu$ is given by $\lambda_{\nu\f}=\la_\f^0/[\sqrt{\tver}-(-1)^{\nu}\sqrt{\tver-1}]$.
 In Fig. \ref{fig:conductivity-isotrop-x-single} (right panel), we have also plotted the conductivity of the system versus $\tilde{\xi}$ for different strengths of SOC. 
 Similar to the two-band case, by increasing $\tilde{\xi}$ the conductivity decreases rapidly, becomes minimum at  $\tilde{\xi}_{min}\sim\la_\f^{av}/2\pi$, where $\la_\f^{av}=\la_\f^0/\sqrt{\tver}$, 
 and then increases gradually.
 In contrast to the two-band case the conductivity decreases by decreasing the SOC and goes toward zero in the limit of $\tver\rightarrow 1$.
 Actually, by decreasing SOC number of electrons with the same $k$ ($k_{1\f}\approx k_{2\f}$) but opposite velocities are almost equal and consequently conductivity is zero.
  In contrast, above the BCP the velocities in the bands $+$ and $-$ are always in the same directions and the system possesses a non zero conductivity.
 \subsection{Anisotropic case $(\theta\neq 0)$} 
\begin{figure}
  \centerline{\includegraphics[width=48mm]{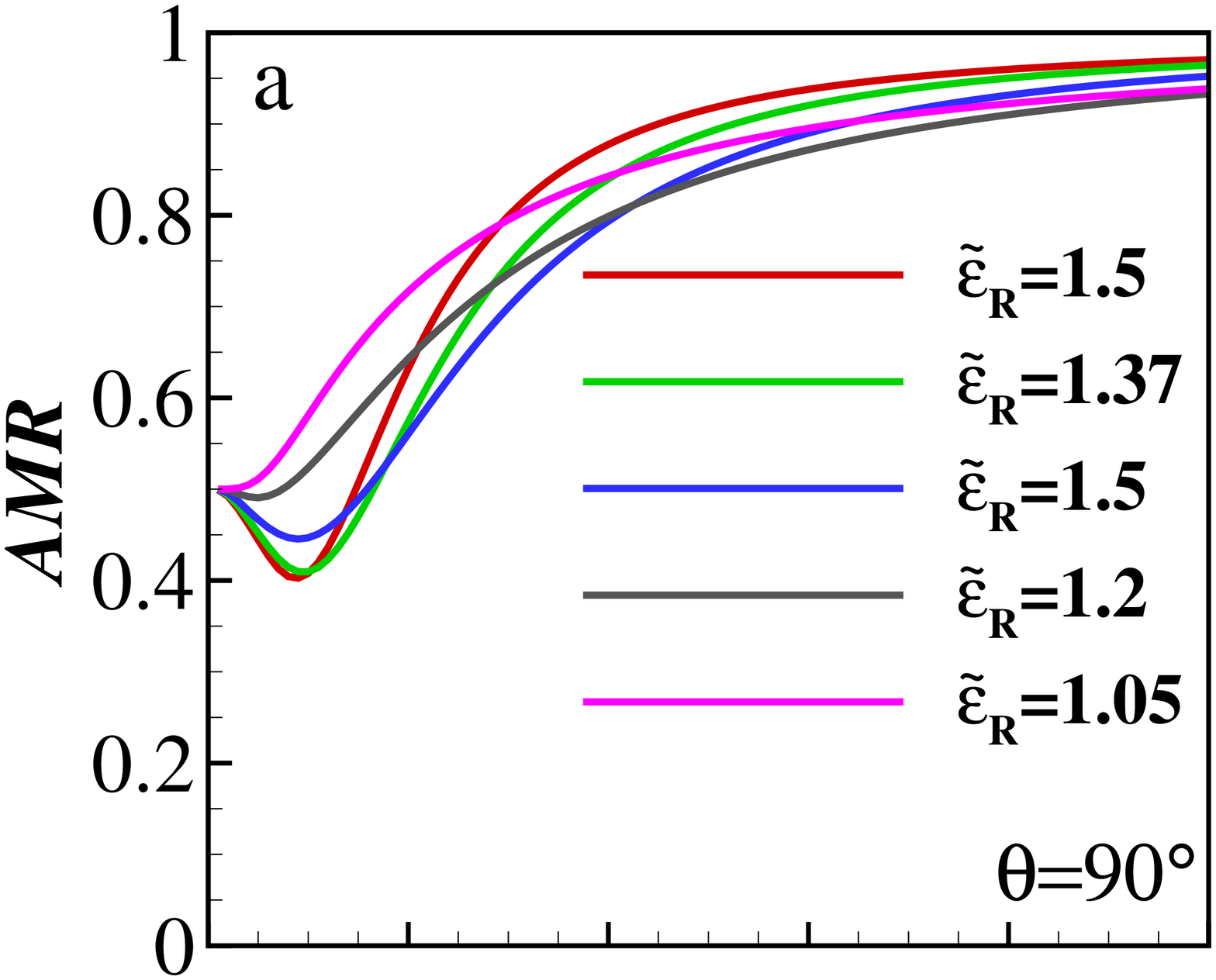}\hspace{-2mm}
 \includegraphics[width=48mm]{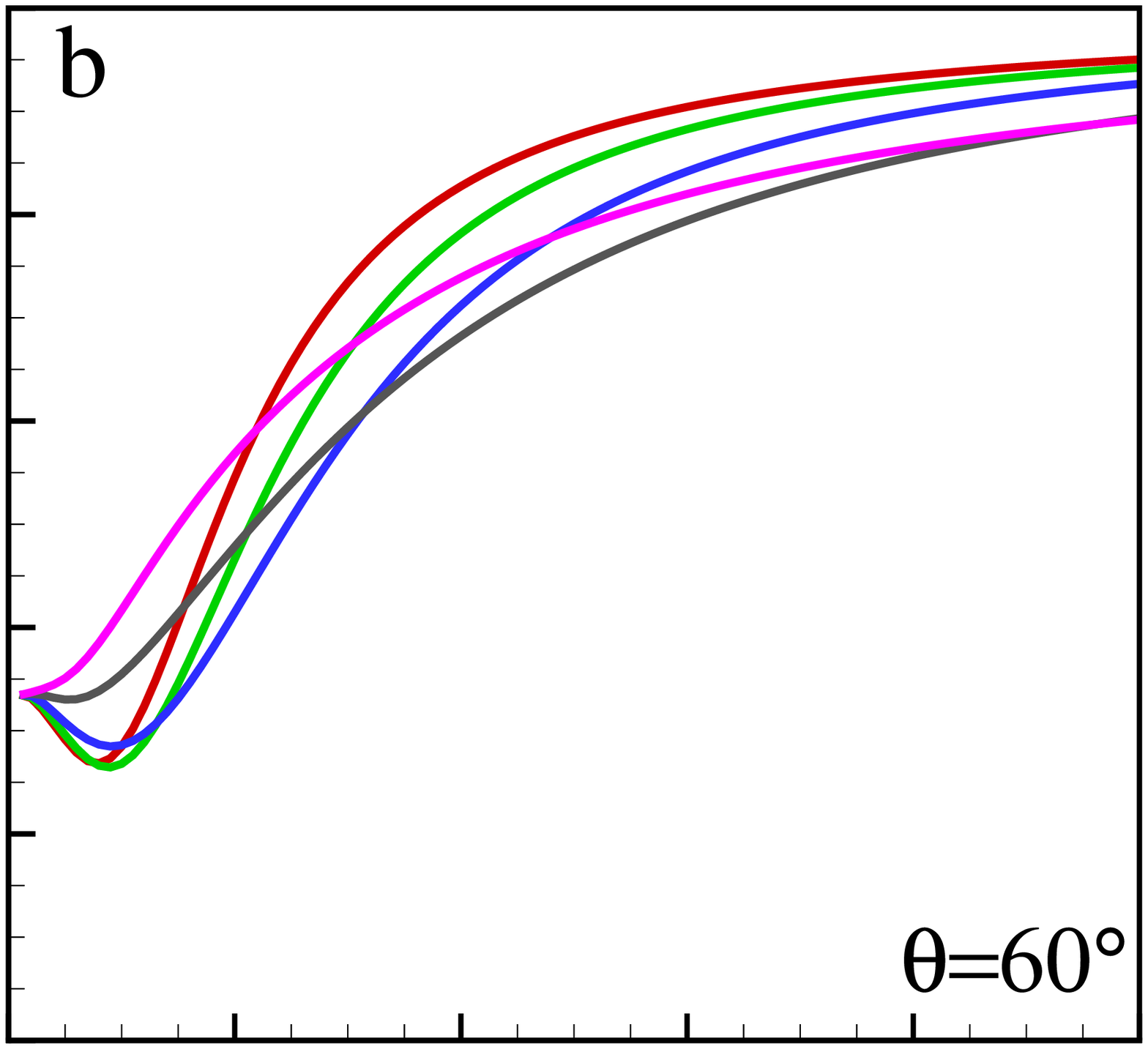}}
 \centerline{\includegraphics[width=48mm]{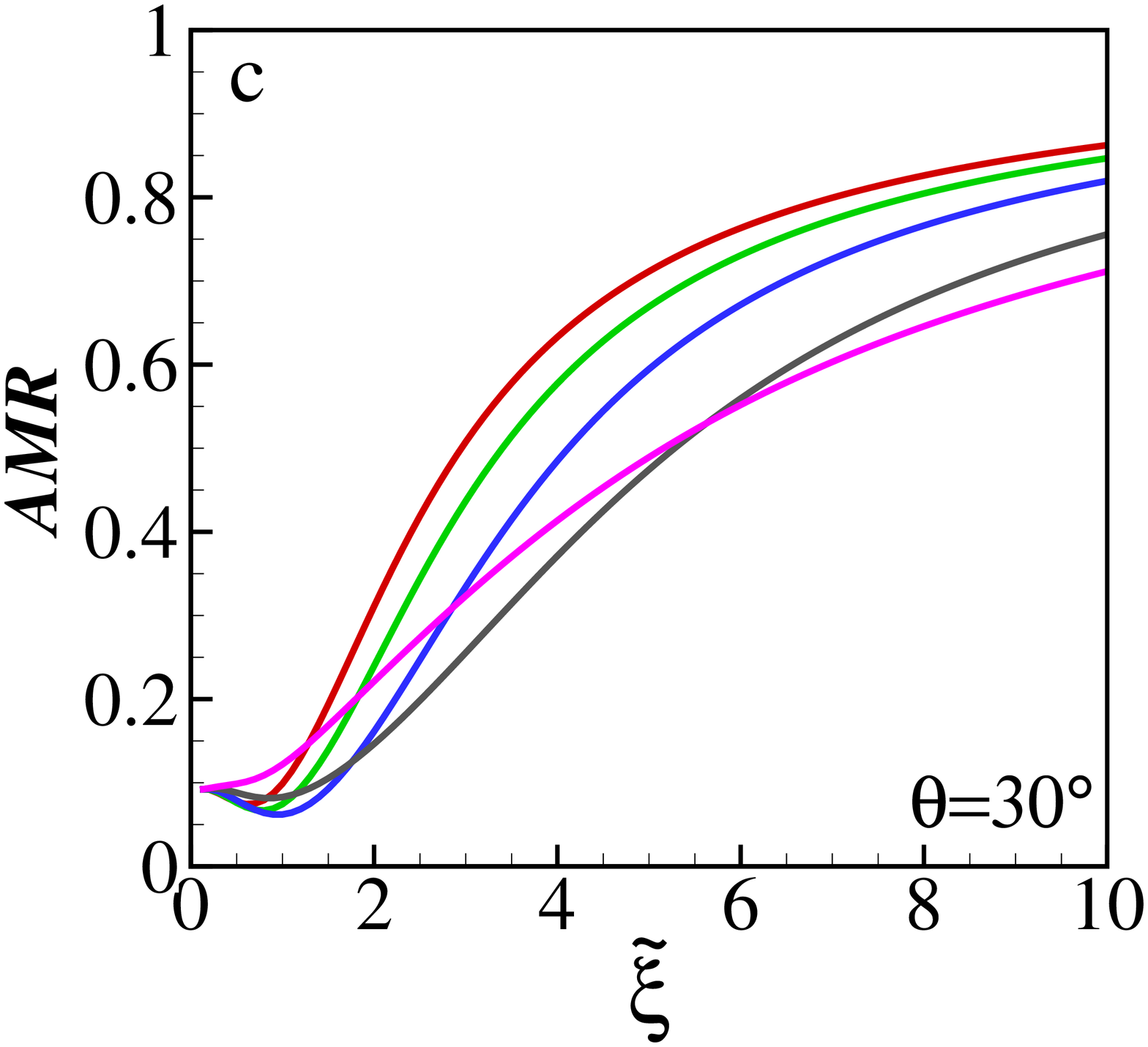}\hspace{-2mm}
 \includegraphics[width=48mm]{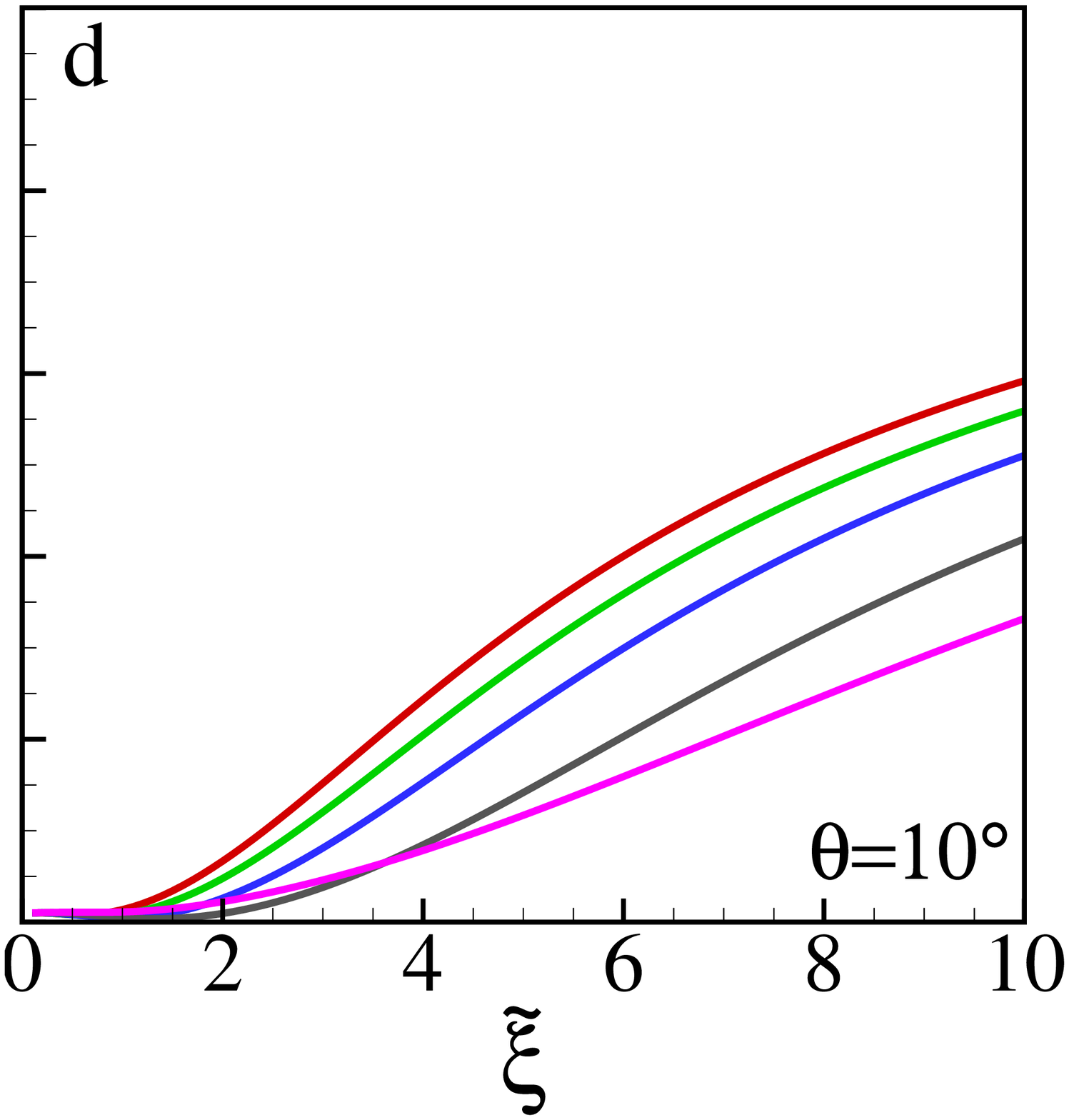}}
 \caption{(Color online) The AMR vs CMS for different strengths of SOC.} \label{fig:amr-single-band}
  \end{figure}

\begin{figure}
                  \centerline{\includegraphics[width=48mm]{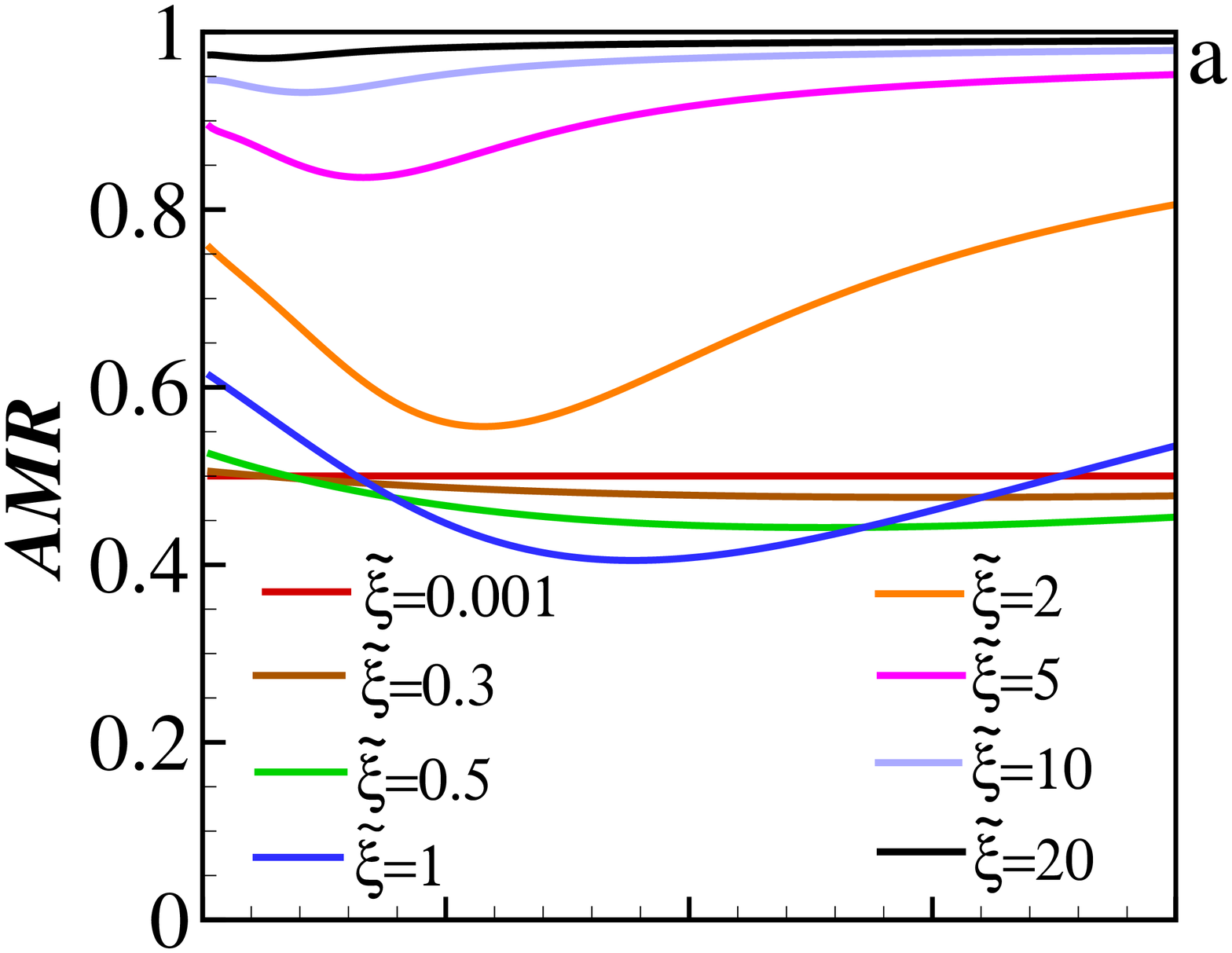}\hspace{-2mm}
                 \includegraphics[width=48mm]{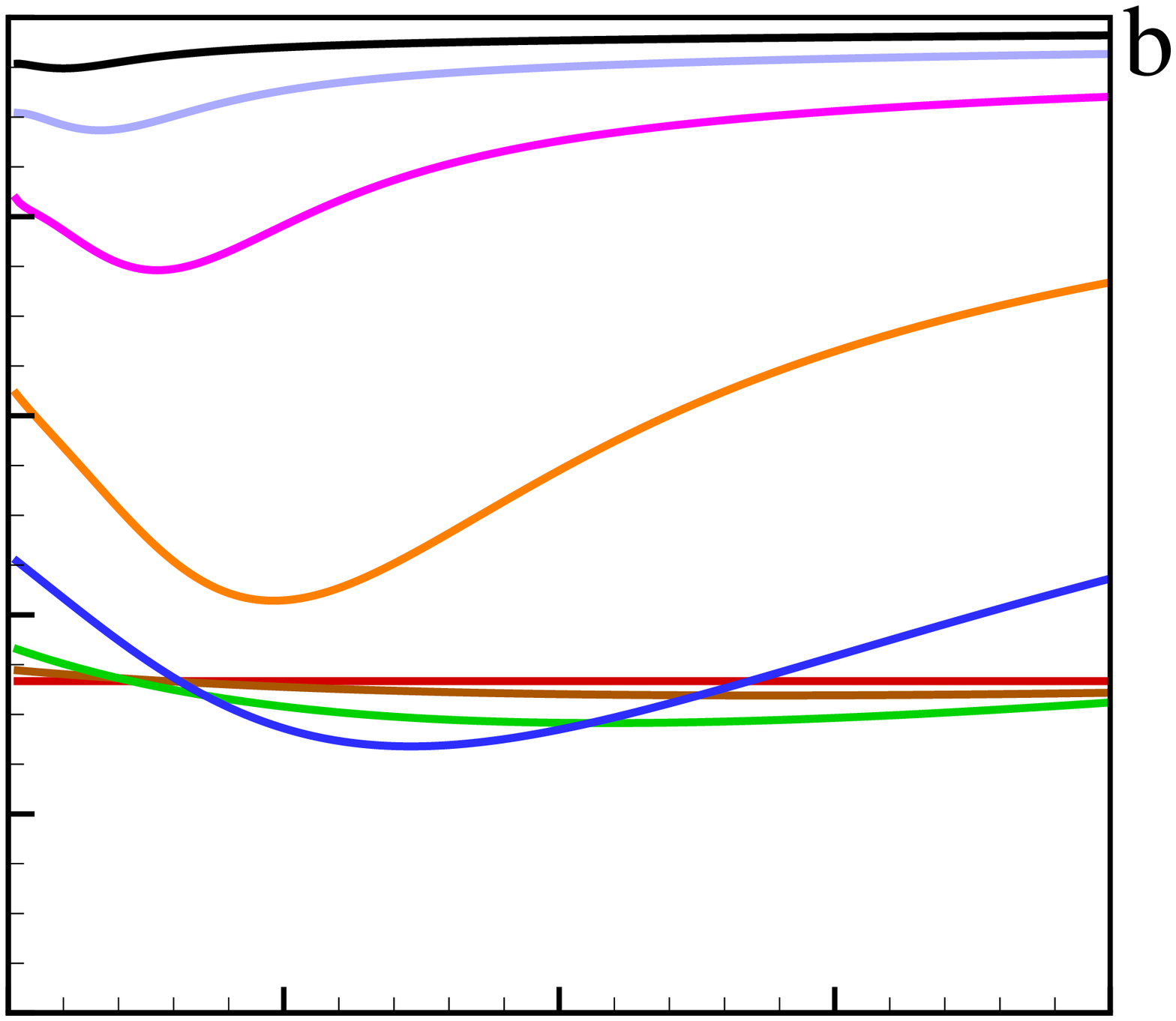}}
                 \centerline{\includegraphics[width=48mm]{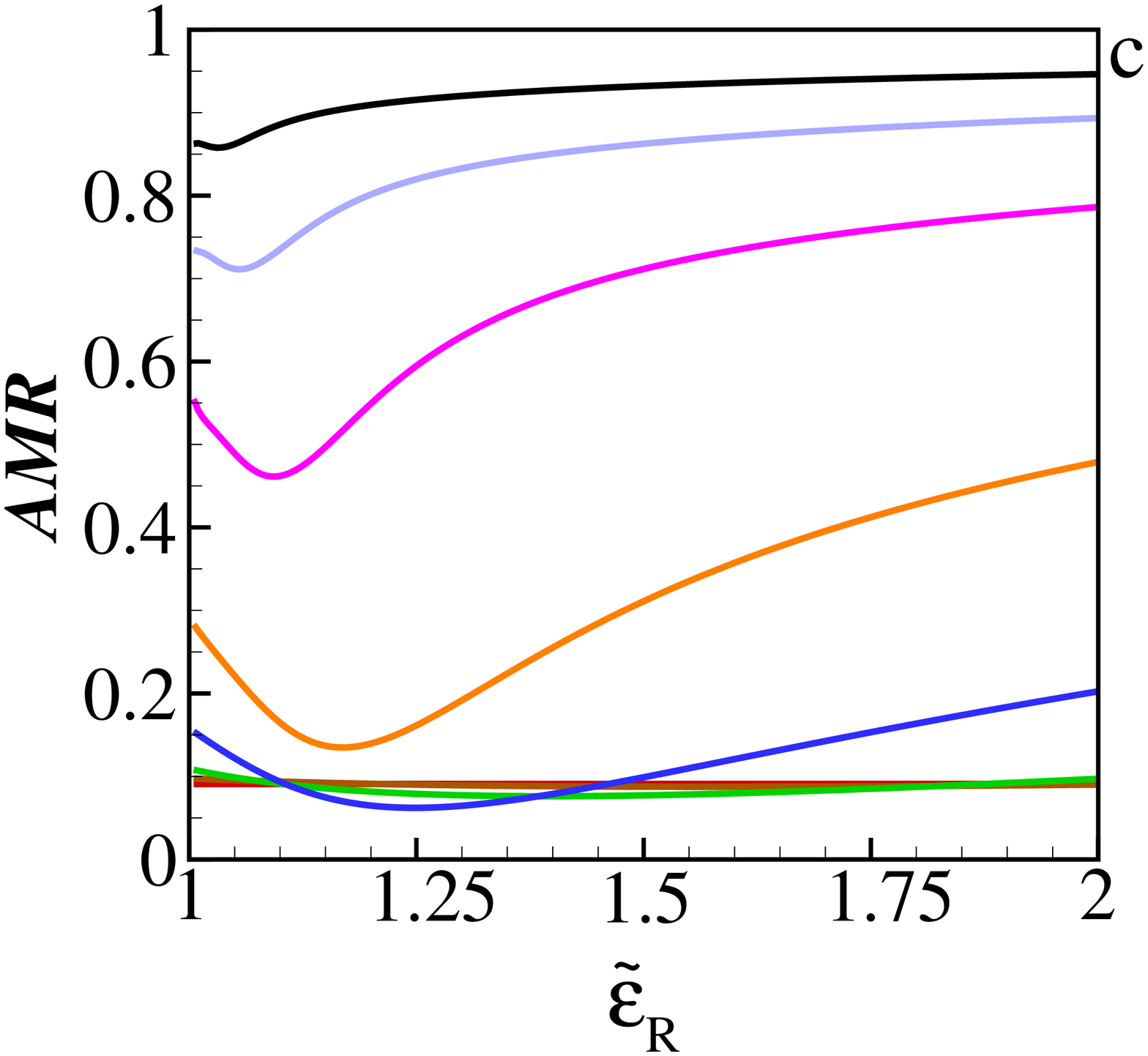}\hspace{-2mm}
                 \includegraphics[width=48mm]{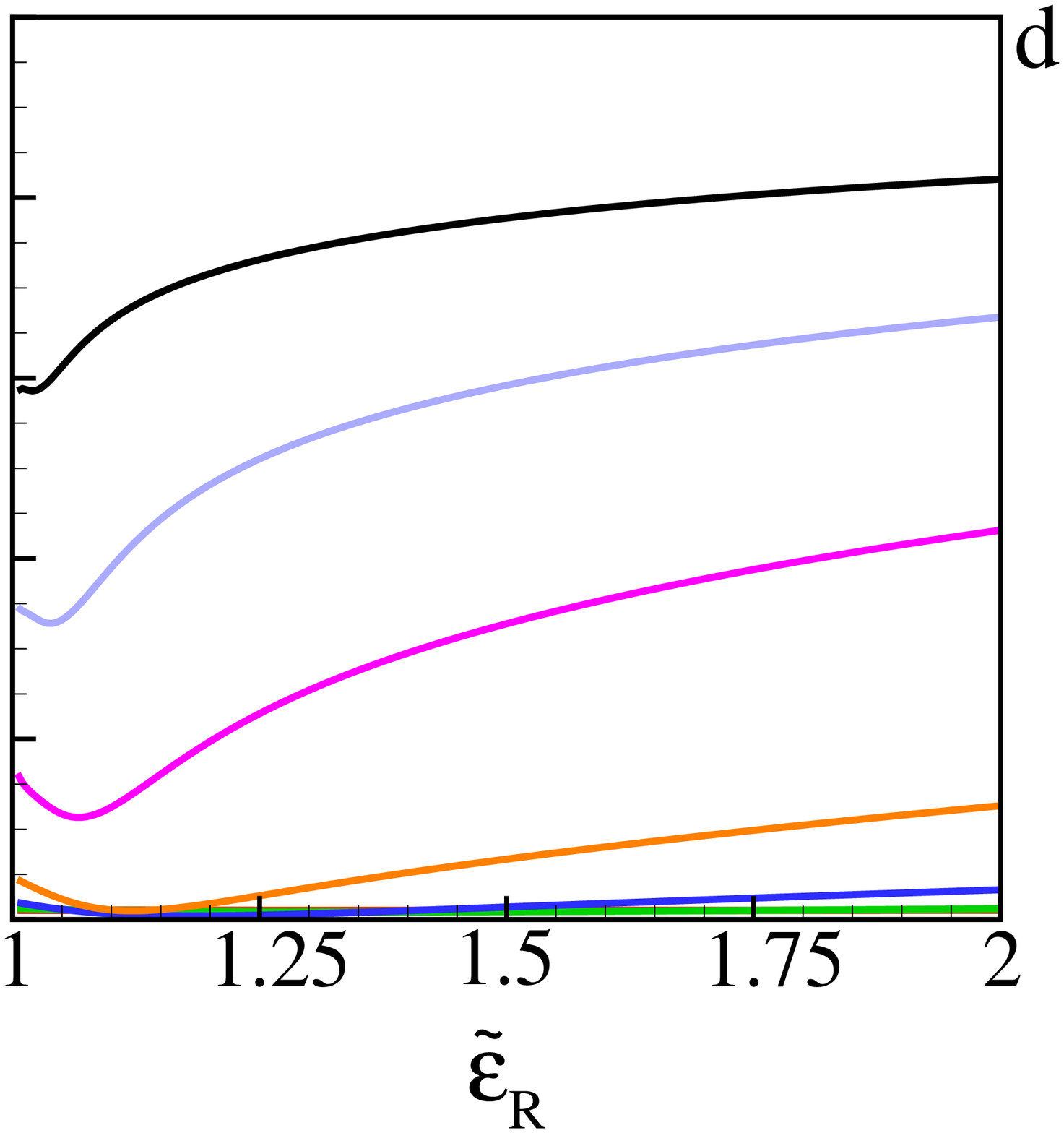}}
                 \caption{(Color online) The AMR versus the strength of SOC, for different CMSs, when a) $\theta=90^\circ$, b)
                 	$\theta=60^\circ$, c) $\theta=30^\circ$ and d) $\theta=10^\circ$.} \label{fig:amr-singleband-alpha}
                  \end{figure}

 In the anisotropic case, below the BCP the behavior of the conductivities $\sigma_{xx}$ and $\sigma_{yy}$, are generally the same as the corresponding ones above the BCP. However, there are some obvious contrasts between these two regimes, which are seen in the behavior of AMR.
 In Fig. \ref{fig:amr-single-band}, we have plotted the AMR of the system versus $\tilde{\xi}$, for different strengths of SOC and various tilt angles $\theta$. Unlike the two-band case, for a given $\theta$ the starting point of the AMR at $\tilde{\xi}=0$ is independent of SOC.
 The behavior of the AMR with respect to CMS is not monotonic for different strengths of SOC. In order to demonstrate such a non-monotonic behavior, we have also plotted in
 Fig. \ref{fig:amr-singleband-alpha}, the AMR versus $\tver$ for different values of $\tilde{\xi}$ and various tilt angles $\theta$. For small CMSs, unlike the two-band case, the AMR is almost independent of $\tver$ and the SOC does not have significant effects on the anisotropy of the system.
  By increasing CMS the AMR becomes more sensitive to $\tver$, it decreases by increasing $\tver$, becomes minimum and then increases to a value less than 1 (this value depends on the CMS). When CMS increases the minimum approaches to the point $\tver=1$. For larger CMSs this minimum disappears and independent of the strength of SOC the AMR becomes almost 1. 
  The reason behind such kind of behavior can be understood by looking at the direction of electric current in the presence of an electric field. When a field is applied along $x$ axis, two types of current are generated in the system: one is by electrons with wave vector $\kv_{1\f}$ along $x$ axis, and the other is by electrons with wave vector $\kv_{2\f}$ in $-x$ direction. 
  Since the current in the branch $1$ is larger than $2$, a net current flows along $x$ direction. Moreover, the Fermi wavelength $\lambda_{1\f}$ is smaller than $\lambda_{2\f}$ for all values of $\tver>1$, therefore for a given $\tilde{\xi}$, by varying $\tver$ electrons with $k_{1\f}$ reaches the efficient scattering point where $\lambda_{1\f}\sim2\pi\xi$, and due to the maximum scattering of electrons in the branch 1 and also the negative conductivity of electrons in the branch 2, $\sigma_{xx}$ is minimum. By increasing $\tver$, Fermi wavelength of electrons in the branch 1 decreases, and positive conductivity along $x$ axis begins to enhance. These variations in $\sigma_{xx}$ cause a minimum to be emerged in the AMR versus $\tver$. 
 By applying the electric field along $y$ axis, the variations of $\sigma_{yy}$ are exactly the same as $\sigma_{xx}$, but since $\sigma_{xx}$ is larger than $\sigma_{yy}$, it has dominant contribution on the AMR.
  By decreasing $\tver$ Fermi wavelength of electrons in the branch 1 increases and hence the minimum approaches to the point $\tilde{\xi}=0$. Also, for large enough CMSs, the Fermi wavelength of electrons in the branch 1 remains always smaller than $\tilde{\xi}$ and no efficient scattering occurs for electrons in this branch, so by increasing $\tilde{\xi}$ this minimum disappears and for very large values of $\tilde{\xi}$ the AMR saturates to 1.
For large values of $\tver$, the Fermi wavelength of electrons in the branch 1 becomes smaller, and a decrease of the CMS causes the minimum location to approach to infinite and the AMR to become a constant (this constant is determined by the title angle of cluster's spin). The main reason for this behavior is the anisotropic band structure below the BCP.
                  \begin{figure}[h]
                  	\centerline{\includegraphics[width=70mm]{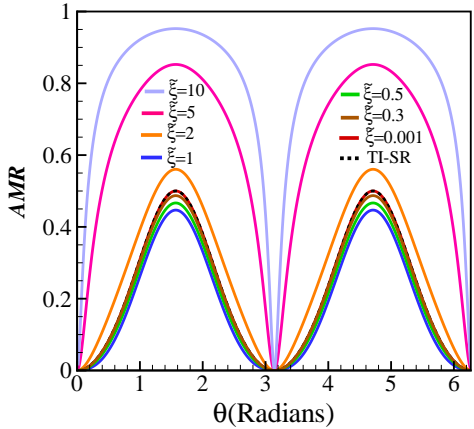}}
                  	\caption{(Color online) The AMR as a function of the tilt angle $\theta$, for different CMSs, when $\tver=1.25$. The black dashed line is for TIs doped with magnetic impurities, where the scattering potential is Dirac $\delta$ function.} \label{fig:amr-teta-single-band}
                  \end{figure}     
  
Fig. \ref{fig:amr-teta-single-band} shows the angular dependence of the AMR. Like the two-band case, the AMR behaves unconventionally for large CMSs (for $\tilde{\xi}$ almost larger than 1). 
For small CMSs the AMR is given by
     \begin{equation}
     AMR=\sin^2\theta/(2+\cos^2\theta),
     \end{equation} 
     which is nothing but the AMR of 3D magnetic topological insulators with short-range  Dirac $\delta$ scattering potential.\cite{0953-8984-27-11-115301,PhysRevB.98.155413}             
 
 \section{Summary and conclusion}\label{sec:conclusion}
 
We presented a comprehensive study of the transport properties of 2DRSs with strong SOC doped with magnetic impurities interacting via an exchange interaction. Exchange interactions between magnetic impurities cause the formation of magnetic clusters which their mean sizes (CMSs) and number (CN) depend on temperature. Treating magnetic clusters as scattering centers, distributed randomly on the surface of the 2DRS, we modeled the interaction of itinerant electrons with magnetic clusters by a long-range scattering potential, and demonstrated that the combined effects of Rashba SOC and magnetic clusters cause the system to be anisotropic which the anisotropy strongly depends on both the clusters' mean size and spin, the strength of SOC and the location of Fermi energy with respect to the BCP.
Using semi-classical Boltzmann approach we computed the relaxation times, the conductivities and the AMRs of the system in both regimes of above and below the BCP. We demonstrated that: 
(i) In the isotropic case the conductivity is a non-monotonic function of CMS, it decreases by increasing CMS, becomes minimum at $\tilde{\xi}_{min.}$, and then increases by CMS. By increasing the strength of SOC the conductivity increases for all CMSs. 
(ii) In the anisotropic case, the AMR strongly depends on the CMS, it increases by increasing CMS and saturates to unit at large CMSs. Moreover, the angular dependence of the AMR is unconventional in comparison with the well-known $\cos^2\theta$ angular
dependence seen in ferromagnets. For small CMSs, below the BCP the angular dependence of the AMR is consistent with 3D magnetic topological insulators.
(iii) In contrast to the two-band regime in which the AMR always increases by increasing the strengths of SOC, in the single band regime it experiences a minimum at a $\tver$ which strongly depends on the CMS.  For small CMSs, in the single band regime the AMR is almost constant and does not change by varying $\tver$, however in the two-band regime the AMR strongly depends on the strength of SOC even at small CMSs.

Both the CMS and CN depend on temperature. By knowing their temperature dependence, the temperature dependence of conductivities can be obtained which is crucial in investigation of the thermoelectric properties of the 2DRSs. The temperature dependences of the CMS and CN can be obtained using Monte Carlo simulations, which are left for future study.




\begin{widetext}

\appendix

\section{The conductivity of 2DRS above the BCP}\label{Appendix-two-band}

When Fermi energy is located above the BCP, both bands are involved in the transport properties of the 2DRS, and we should take the contributions of backscatteringss and intra-bands scatterings into account in computing the non-equilibrium distribution functions ($f_+$ and $f_-$).  
Substituting Eq. (\ref{eq:relax-time-anisotropic}) into (\ref{eq:boltzmann2}), and writing ${\bf E}\cdot\v_n$ as $Ev_n\cos(\phi-\chi)$, we achieve the following integral equations,
\begin{equation}
\begin{aligned}
& \cos\phi=\bar{w}_{+}(\phi)a_{+}(\phi)-\int d\phi{'}[w_{++}(\phi,\phi{'})a_{+}(\phi{'})\\
&+w_{+-}(\phi,\phi{'})a_{-}(\phi{'})],\\
&\cos\phi=\bar{w}_{-}(\phi)a_{-}(\phi)-\int d\phi{'}[w_{--}(\phi,\phi{'})a_{-}(\phi{'})\\
&+w_{-+}(\phi,\phi{'})a_{+}(\phi{'})],
\end{aligned}
\label{ac+-}
\end{equation} 
\begin{equation}
\begin{aligned}
& \sin\phi=\bar{w}_{+}(\phi)b_{+}(\phi)-\int d\phi{'}[w_{++}(\phi,\phi{'})b_{+}(\phi{'})\\
&+w_{+-}(\phi,\phi{'})b_{-}(\phi{'})],\\
&\sin\phi=\bar{w}_{-}(\phi)b_{-}(\phi)-\int d\phi{'}[w_{--}(\phi,\phi{'})b_{-}(\phi{'})\\
&+w_{-+}(\phi,\phi{'})b_{+}(\phi{'})],
\end{aligned}
\label{bs+-}
\end{equation} 
where $w_{n,n'}(\phi,\phi')=A(2\pi)^{-2}\int k' dk' w_{n,n'}(\kv,\kv')$, and $\bar{w}_{n}(\phi)=\int d\phi'[w_{+n}(\phi,\phi')+w_{-n}(\phi,\phi')]$. 
Using Fermi golden rule, the transition rate is written in terms of the
scattering amplitude $|T_{n,n'}(\kv,\kv')|^2$ as:
\begin{equation}
w_{n,n'}(\kv,\kv')=\frac{2\pi}{\hbar}|T_{n,n'}(\kv,\kv')|^2\delta(\ve_{n k}-\ve_{n'k'}).\label{eq:transition-rate1}
\end{equation}
Within the first Born approximation, the $T$-matrix is given by:
$T_{n,n'}(\kv,\kv')\approx\langle n\kv|{\cal H}_{\sigma S}|n'\kv'\rangle$,
where ${\cal H}_{\sigma S}$ is the scattering Hamiltonian, given by
${\cal H}_{\sigma S}=\sum_{\rv,\Rv}H_{\sigma S}$. Since magnetic
clusters are uncorrelated and distributed randomly on 
the 2DRS, one can show that $\langle|T_{n,n'}(\kv,\kv')|^2\rangle _{ens.}=\eta_{c}|H_{\sigma S}(\kv-\kv')|^{2}$,
where $\langle\dots\rangle_{ens.}$ denotes ensemble average, $\eta_c$ is the number of clusters with mean size (CN), and $H_{\sigma S}(\kv-\kv')$ is the Fourier
transformation of Eq. ( \ref{eq:scattering-Hamiltonian}) at $\Rv=0$. As the unperturbed Hamiltonian in Eq. (\ref{eq:hamiltonian}) is gapless and the time reversal symmetry is preserved in the system, other mechanisms such as skew scattering, anomalous velocity, and side-jump have vanishing contributions to the transport properties of the system (anomalous Hall conductivity is zero) and 
the lowest-order Born approximation gives reliable results for the transport of the system.\cite{Sinitsyn_2007}

By considering elasticity of the scattering we obtain,
\begin{equation}
\begin{aligned}
& w_{n,n'}= \omega_0\frac{N_n}{N_0}\tilde{\xi}^4\frac{1-n n'(\cos 2\theta\cos\phi\cos\phi'
	+\sin\phi\sin\phi')}{(1+\xi^2[k^2_n+k^2_{n'}])^3(1-\Omega_{n,n'}\cos\Delta\phi)^{3}}, \\
& \bar{w}_n(\phi)=\omega_0\sum_{n'=\pm}\frac{\tilde{\xi}^4\left(1+n'\sqrt{\frac{\ve_\r/2\ve}{\ve_\r/2\ve+1}}\right)}{(1+\xi^{2}[k^{2}_n+k^{2}_{n'}])^{3}(1-\Omega_{n,n'}^{2})^{\frac{5}{2}}}\times\\
&\left[\frac{3}{2}\pi n n' \Omega_{n,n'}\sin^{2}\theta\cos2\phi+\pi (2+\Omega_{n,n'}^2-3 n n'\cos^{2}\theta)\right], 
\label{two-band-w}
\end{aligned}
\end{equation}
where $\Omega_{n,n'}=\frac{2\xi^2 k_n k_{n'}}{1+\xi^2[k_n^2+k_{n'}^2]}$, $\omega_{0}=\frac{\pi \hbar \eta_c J_{0}^{2}S^2 }{4 A m\varepsilon_\f}$, and $\tilde{\xi}=\xi\sqrt{2m\ve_\f/\hbar^2}$.

In order to solve Eqs. (\ref{ac+-}) and (\ref{bs+-}), we employ the Fourier expansions of $a_{\pm}(\phi)$ and $b_{\pm}(\phi)$ as
\begin{eqnarray}
a_{\pm}(\phi)&=&\sum_{m=0}^{\infty}c^{\pm}_{2m+1}\cos[(2m+1)\phi],
\label{reduced-a}\\
b_{\pm}(\phi)&=&\sum_{m=0}^{\infty}s^{\pm}_{2m+1}\sin[(2m+1)\phi],
\label{reduced-b}
\end{eqnarray}
which satisfy the particle number conservation.
Since the functions  $w_{n,n'}(\phi,\phi')$ and $\bar{w}_{n}(\phi)$ are invariant under the transformations $(\phi,\phi^{'})\rightarrow (-\phi,-\phi^{'})$ and $(\phi,\phi^{'})\rightarrow (\pi-\phi,\pi-\phi^{'})$, the functions $a_{\pm}(\phi)$ and $b_{\pm}(\phi)$ should satisfy the relations: $a_{\pm}(-\phi)=a_{\pm}(\phi)$,  $a_{\pm}(\pi-\phi)=-a_{\pm}(\phi)$,  $b_{\pm}(-\phi)=-b_{\pm}(\phi)$, and  $b_{\pm}(\pi-\phi)=b_{\pm}(\phi)$, hence only the Fourier coefficients $c_i^\pm$ and $s_i^\pm$ with odd $i$ has been appeared in the above expansions. $c_{2m+1}^\pm$ and $s_{2m+1}^\pm$ have a dimension of time and depend on the CMS, the CN, the tilt angle $\theta$, and the Rashba energy $\ve_\r$. By substituting Eq. (\ref{reduced-a}) into the both relations in Eq. (\ref{ac+-}), we obtain the following set of linear equations for the coefficients $c^{\pm}_{i}$,
\begin{equation}
\begin{bmatrix} {\bf A}_{0}&{\bf C}_{1}&{\bf 0}&\dots&\dots&\dots\\
{\bf C}_{1}&{\bf A}_{1}&{\bf C}_{2}&{\bf 0}&\dots&\dots&\dots\\
{\bf 0}&{\bf C}_{2}&{\bf A}_{2}&{\bf C}_{3}&{\bf 0}&\dots&\dots\\
{\bf 0}&{\bf 0}&{\bf C}_{3}&{\bf A}_{3}&{\bf C}_{4}&{\bf 0}&\dots\\
{\bf 0}&{\bf 0}&{\bf 0}&{\bf C}_{4}&{\bf A}_{4}&{\bf C}_{5}&\dots\\
\vdots&\vdots&\vdots&\vdots&\vdots&\ddots
\end{bmatrix}
\begin{bmatrix}\c_1\\
\c_3\\
\c_5\\
\c_7\\
\c_9\\
\vdots
\end{bmatrix}
=1/\omega_0
\begin{bmatrix}\d_1\\
\d_3\\
\d_5\\
\d_7\\
\d_9\\
\vdots
\end{bmatrix}, 
\label{mc}
\end{equation}
where the matrix elements are the following $2\times 2$ matrices;  
\begin{equation}
{\bf A}_{0}  =\begin{bmatrix} L_{0}^{+}+K_{0}^{+} & G_{0}^{-}+D_{0}^{-}\\
G_{0}^{+}+D_{0}^{+} & L_{0}^{-}+K_{0}^{-}       
\end{bmatrix},
\label{Eq:b0}
\end{equation} 
and
\begin{equation}
{\bf A}_{m}  =\begin{bmatrix} K_{m}^{+} & D_{m}^{-}\\
D_{m}^{+} & K_{m}^{-}       
\end{bmatrix},~~~~~ {\bf C}_{m}  =\begin{bmatrix} L_{m}^{+} & G_{m}^{-}\\
G_{m}^{+} & L_{m}^{-}      
\end{bmatrix},
\label{Eq:bn-cn}
\end{equation}  
where $L_{m}^{\pm}$, $K_{m}^{\pm}$, $G_{m}^{\pm}$ and $D_{m}^{\pm}$ are the following dimensionless functions:
\begin{equation}
\begin{aligned}
L_{m}^{\pm}(\varepsilon,\xi,\theta)/\tilde{\xi}^4&=\kappa_{\pm\pm}+\kappa_{\pm\mp}-\frac{F(m,\Omega_{\pm\pm})(1-\cos2\theta)(1\pm\sqrt{\frac{\ve_\r/2\ve}{\ve_\r/2\ve+1}})}{(1+2\xi^{2}k_\pm^2)^{3}},\\
K_{m}^{\pm}(\varepsilon,\xi,\theta)/\tilde{\xi}^4&=\mu_{\pm\pm}+\mu_{\pm\mp}-\frac{\Big[Q(m,\Omega_{\pm\pm})-\Big\{F(m,\Omega_{\pm\pm})+F(m+1,\Omega_{\pm\pm})\Big\}(1+\cos2\theta)\Big](1\pm\sqrt{\frac{\ve_\r/2\ve}{\ve_\r/2\ve+1}})}{(1+2\xi^2k_\pm^2)^3},\\
G_{m}^{\pm}(\varepsilon,\xi,\theta)/\tilde{\xi}^4&=\frac{F(m,\Omega_{\pm\mp})(1-\cos2\theta)(1\pm\sqrt{\frac{\ve_\r/2\ve}{\ve_\r/2\ve+1}})}{(1+\xi^2[k_\pm^2+k_\mp^2])^{3}},\\
D_{m}^{\pm}(\varepsilon,\xi,\theta)/\tilde{\xi}^4&=-\frac{\Big[Q(m,\Omega_{\pm\mp})+\Big\{F(m,\Omega_{\pm\mp})+F(m+1,\Omega_{\pm\mp})\Big\}(1+\cos2\theta)\Big](1\pm\sqrt{\frac{\ve_\r/2\ve}{\ve_\r/2\ve+1}})}{(1+\xi^2[k_\pm^2+k_\mp^2])^{3}},\\
\label{eq:Lm-Km}
\end{aligned}
\end{equation}
where
\begin{equation}
\kappa_{nn'}=\frac{\frac{3}{2}\pi n n' \Omega_{n,n'}\sin^{2}\theta(1+n'\sqrt{\frac{\ve_\r/2\ve}{\ve_\r/2\ve+1}})}{(1+\xi^{2}[k^{2}_n+k^{2}_{n'}])^{3}(1-\Omega_{n,n'}^{2})^{\frac{5}{2}}},
\end{equation}
\begin{equation}
\mu_{nn'}=\frac{\pi (2+\Omega_{n,n'}^2-3 n n'\cos^{2}\theta)(1+n'\sqrt{\frac{\ve_\r/2\ve}{\ve_\r/2\ve+1}})}{(1+\xi^{2}[k^{2}_n+k^{2}_{n'}])^{3}(1-\Omega_{n,n'}^{2})^{\frac{5}{2}}},
\end{equation}
and
\begin{equation}
\begin{aligned}
F(m,y)=\sum_{l=0}\frac{\pi(2l+2)!}{4(l-m)!(l+m)!}\left(\frac{y}{2}\right)^{2l}
=&\sum_{k=0}\frac{\pi(2m+2k+2)!}{4k!(2m+k)!}\left(\frac{y}{2}\right)^{2m+2k}\\[5pt]
=&\frac{\pi(1+m)(1+2m)}{2}\left(\frac{y}{2}\right)^{2m}{}_2F_{1}[\frac{3}{2}+m,2+m,1+2m,y^{2}],\\[5pt]
Q(m,y)=\sum_{l=0}\frac{\pi(2l+3)!}{(l-m)!(l+m+1)!}\left(\frac{y}{2}\right)^{2l+1}
=&\sum_{k=0}\frac{\pi(2m+2k+3)!}{k!(2m+k+1)!}\left(\frac{y}{2}\right)^{2m+2k+1}\\[5pt]
=&2\pi(m+1)(3+2m)\left(\frac{y}{2}\right)^{2m+1}{}_2F_{1}[2+m,\frac{5}{2}+m,2+2m,y^{2}].
\end{aligned}
\label{eq:hyper}
\end{equation}
Here, ${}_2F_{1}[\frac 32+m,2+m,1+2m,y^2]$ and
${}_2F_{1}[2+m,\frac 52+m,2+2m,y^2]$ are hypergeometric
functions. Since for $\ve_\r/\ve_\f\geq 0$, we have $0\leq\Omega_{n,n'}<1$, the functions $F(m,\Omega_{n,n'})$ and $Q(m,\Omega_{n,n'})$ are simplified as
\begin{equation}
\begin{aligned}
&F(m,\Omega_{n,n'})=\frac{\pi(2+\Omega_{n,n'}^{2}+6m\sqrt{1-\Omega_{n,n'}^{2}}-4m^{2}(1-\Omega_{n,n'}^{2}))}{4 \Omega_{n,n'}^{-2m}(1+\sqrt{1-\Omega_{n,n'}^{2}})^{2m}(1-\Omega_{n,n'}^{2})^{\frac{5}{2}}},\\[5pt]
& Q(m-1,\Omega_{n,n'})=\frac{3\pi-3\pi\sqrt{1-\Omega_{n,n'}^{2}}(1-2m)+4\pi m(m-1)(1-\Omega_{n,n'}^{2})}{\Omega_{n,n'}^{1-2m}(1+\sqrt{1-\Omega_{n,n'}^{2}})^{2m-1}(1-\Omega_{n,n'}^{2})^{\frac{5}{2}}}.
\end{aligned}
\end{equation}
 The vectors $\c_m$ and $\d_m$ in Eq. (\ref{mc}) are the following two-component vectors:
\begin{equation}
\begin{aligned}
& \c_m= \begin{bmatrix} c^{+}_m\\
c^-_m
\end{bmatrix},
&\d_m= \begin{cases}
\begin{bmatrix} 1\\
1 
\end{bmatrix}, & \text{ $m> 1$ } \\
{\bf 0} & \text{$m=1$}.
\end{cases}
\end{aligned}
\end{equation} 
For solving the linear equations in (\ref{mc}), we have to truncate the series (\ref{reduced-a}) at some point. For an specified $\xi$, $\theta$ and $\ve_\r$, by selecting a proper number of trigonometric functions in the series (\ref{reduced-a}), we obtain $a_{+}(\phi)$ and $a_{-}(\phi)$ functions, precisely.\cite{PhysRevB.98.155413} By selecting the first $j$ independent trigonometric functions, the matrix equation (\ref{mc}) reduces to
\begin{equation}
\begin{bmatrix} {\bf A}_{0}& {\bf C}_{1}& {\bf 0}&{\bf 0}& \dots& {\bf 0}\\
{\bf C}_{1}&{\bf A}_{1}& {\bf C}_{2} & {\bf 0} & \dots & {\bf 0}\\
{\bf 0}& {\bf C}_{2} & {\bf A}_{2} & {\bf C}_{3} & \dots & \vdots\\
\vdots&\vdots&\vdots&\vdots&\vdots& {\bf 0}\\
{\bf 0} & \dots & {\bf 0}& {\bf C}_{j-1} & {\bf A}_{j-1}& {\bf C}_j\\
{\bf 0} & \dots & {\bf 0} & {\bf 0} & {\bf C}_j & {\bf A}_j
\end{bmatrix}
\begin{bmatrix}\c_1\\
\c_3\\
\c_5\\
\vdots \\
\c_{2j-1}\\
\c_{2j+1}\\
\end{bmatrix}
=
1/\omega_{0} \begin{bmatrix}\d_1\\
\d_3\\
\d_5\\
\vdots\\
\d_{2j-1}\\
\d_{2j+1}\\
\end{bmatrix}.
\label{cuted-ma}
\end{equation}
Now we have a set of $j+1$ equations with $j+1$ variables. Since $\d_{2j+1}={\bf 0}$ for $j\geq1$, we begin with the two last equations $j+1$ and $j$, then by eliminating the coefficient $\c_{2j+1}$ we reach to the following equation
\begin{equation}
{\bf C}_{j}^{-1}\cdot {\bf C}_{j-1}\cdot \c_{(2(j-2)+1)}+\Delta_{j}\cdot \c_{(2(j-1)+1)}=0,
\label{jth}
\end{equation}
where
\begin{equation}
\Delta_{j}={\bf C}_{j}^{-1}\cdot {\bf A}_{j-1}-{\bf A}_{j}^{-1}\cdot {\bf C}_{j}.
\label{dl}
\end{equation}  
By using the $(j-1)$th equation of (\ref{cuted-ma}) and Eq. (\ref{jth}) we can eliminate the coefficient $\c_{2j-1}$. Repeating this method to the first equation of (\ref{cuted-ma}) we reach to        
\begin{equation}
\omega_0 \c_1=
(\Delta^{+}_{1})^{-1}\cdot {\bf C}_{1}^{-1}\cdot \d_1
\label{ac1},
\end{equation} 
where,
\begin{equation}
\Delta^{+}_{i}={\bf C}_{i}^{-1}\cdot {\bf A}_{i-1}-(\Delta^{+}_{i+1})^{-1}\cdot {\bf C}_{i+1}^{-1}\cdot {\bf C}_{i}~.
\end{equation}   
By obtaining $\c_1$ from Eq. (\ref{ac1}) and substituting it into Eq. (\ref{mc}), the other $\c_i$ vectors, and finally the functions $a_{\pm}(\phi)$ are obtained in terms of $k, \alpha, \xi, \theta$, and $\eta_c$.

By the same procedure, we have also obtained the functions $ b_\pm(\phi)$ in terms of the non-zero coefficients $s_i^\pm=s_{i}^{ \pm}(k, \alpha, \xi, \theta, \eta_c)$, with odd $i$. All these nonzero coefficients are given in terms of ${\bf s}_1$ by the relation
\begin{equation}
\omega_0  {\bf s}_1=
(\Delta^{-}_{1})^{-1}\cdot {\bf C}_{1}^{-1}\cdot \d_1,
\label{b1}
\end{equation} 
where ${\bf s}_m={\begin{bmatrix} s^{+}_{m}\\
	s^{-}_{m}
	\end{bmatrix}}$, and
\begin{equation} 
\Delta^{-}_{1}={\bf C}_{1}^{-1}\cdot {\bf B}_{0}-(\Delta^{+}_{2})^{-1}\cdot {\bf C}_{2}^{-1}\cdot {\bf C}_{1},
\end{equation}
with 
\begin{equation}
{\bf B}_{0}  =\begin{bmatrix} K_{0}^{+}-L_{0}^{+}& D_{0}^{-}-G_{0}^{-}\\
D_{0}^{+}-G_{0}^{+} & K_{0}^{-}-L_{0}^{-}       
\end{bmatrix}.
\label{b'0}
\end{equation}    

\section{The conductivity of the 2DRS below the BCP}\label{Appendix-single-band}
When the Fermi energy is located in the interval $-\ve_\r/2<\ve_\f<0$, we have unconventional intra-branches and inter-branches scatterings in the band $+$, with the scattering rate:   
\begin{equation}
\begin{aligned}
& w_{\nu,\nu'}=\omega_0\frac{N_\nu}{N_0}\tilde{\xi}^4\frac{1-\cos2\theta\cos\phi\cos\phi'
	-\sin\phi\sin\phi'}{(1+\xi^2[k_\nu^2+k_{\nu'}^2])^3(1-\Gamma_{\nu,\nu'}\cos\Delta\phi)^3},\\
& \bar{w}_\nu(\phi)=\omega_0\sum_{\nu'}\frac{\tilde{\xi}^4\left((-1)^{1+\nu-\nu'}+\sqrt{\frac{\ve_\r/2|\ve|}{\ve_\r/2|\ve|-1}}\right)}{(1+\xi^{2}[k_{\nu}^2 +k_{\nu'}^2 ])^{3}(1-\Gamma_{\nu,\nu'}^{2})^{\frac{5}{2}}}\times\\
&~~~\left[\frac{3}{2}\pi \Gamma_{\nu,\nu'}\sin^{2}\theta\cos2\phi+\pi (2+\Gamma_{\nu,\nu'}^2-3\cos^{2}\theta)\right],
\label{single-band-omega}
\end{aligned}    
\end{equation}
with $\Gamma_{\nu,\nu'}=\frac{2\xi^{2}k_\nu k_{\nu'}}{1+\xi^{2}[k_{\nu}^{2}+k_{\nu'}^{2}]}$.  
In order to obtain the non-equilibrium distribution functions of electrons in the branches 1 and 2, we should solve two equations like (\ref{ac+-}) and (\ref{bs+-}).              
Because of the non-monotonic dispersion of the band $+$, below the BCP the band velocity of electrons is not always parallel to their $\kv$-vector.  In order to include this issue in our calculations, we replace $\phi$ by ${\Phi}_\nu(\phi)$, defined by ${\bf v}_\nu(\ve,\phi)=v_\nu(\ve,\phi)(\cos\Phi_\nu,\sin\Phi_\nu)$, where $\Phi_1=\phi$ and $\Phi_2=\phi+\pi$. Therefore, according to the relation $\v_\nu=(-1)^\nu\frac{\hbar}{m}(N_0/N_\nu)\kv_\nu$, and the method used to obtain the non-equilibrium distribution functions, we define the non-equilibrium distribution functions of electrons in the branch $\nu$ as (see Appendix \ref{A2}): 
\begin{equation}
f_\nu-f^0_\nu=
eE v_\nu\left(\frac{\partial f^0_\nu}{\partial\epsilon_\nu}\right)[a_\nu(\Phi_\nu)\cos\chi+b_\nu(\Phi_\nu)\sin\chi],
\label{f-single-band}
\end{equation} 
where $a_\nu(\Phi_\nu)$ and $b_\nu(\Phi_\nu)$ satisfy the following relations:
\begin{equation}
\begin{aligned}
& \cos\Phi_2=\bar{w}_2(\phi)a_2(\Phi_2)-\int d\phi'[w_{2,2}(\phi,\phi')a_2(\Phi'_2)\\
&+w_{2,1}(\phi,\phi')a_1(\Phi'_1)],\\
&\cos\Phi_1=\bar{w}_1(\phi)a_1(\Phi_1)-\int d\phi'[w_{1,1}(\phi,\phi')a_1(\Phi'_1)\\
&+w_{1,2}(\phi,\phi')a_2(\Phi'_2)],
\end{aligned}
\label{ac12}
\end{equation} 
\begin{equation}
\begin{aligned}
& \sin\Phi_2=\bar{w}_2 (\phi)b_2(\Phi_2)-\int d\phi'[w_{2,2}(\phi,\phi')b_2(\Phi'_2)\\
&+w_{2,1}(\phi,\phi')b_1(\Phi'_1)],\\
&\sin\Phi_1=\bar{w}_1 (\phi)b_1(\Phi_1)-\int d\phi'[w_{1,1}(\phi,\phi')b_1(\Phi'_1)\\
&+w_{1,2}(\phi,\phi')b_2(\Phi'_2)].
\end{aligned}
\label{bs12}
\end{equation} 
For solving the coupled Eqs. (\ref{ac12}) and (\ref{bs12}), the same as the two-band case, we use the Fourier expansions of $a_\nu(\Phi_\nu)$ and $b_\nu(\Phi_\nu)$, and achieve  
\begin{equation}
\begin{aligned}
& a_{\nu}(\Phi_{\nu})=\sum_{m=0}^{\infty}c_{2m+1}^{\nu}\cos[(2m+1)\Phi_{\nu}],\\
& b_{\nu}(\Phi_{\nu})=\sum_{m=0}^{\infty}s_{2m+1}^{\nu}\sin[(2m+1)\Phi_{\nu}],
\end{aligned}
\label{b-a-single-reduced}
\end{equation}
where $c_{2m+1}^{\nu}$ and $s_{2m+1}^{\nu}$ are coefficients with a dimension of time. By using the relation (\ref{conductivity}), and considering the orthogonality of trigonometric functions, the conductivity for $-\frac{\ve_\r}{2}<\ve_\f<0$ is obtained in terms of only the first coefficients of the expansions, $c^\nu_1$ and $s^\nu_1$. As mentioned in the two-band case, these coefficients treat as momentum relaxation times of free electrons and can be regarded as effective relaxation times along $x$ and $y$ directions. By defining  $\t_x^\nu=\omega_0 c_1^\nu$ and
$\t_y^\nu=\omega_0 s_1^\nu$ we obtain the conductivities of the 2DRS, as in Eq. (\ref{sigma-single-band-iso}).

\section{Boltzmann equation below the BCP}\label{A2}

In elastic scatterings, the magnitude of electrons' velocity $v$ does not change, and the non-equilibrium distribution functions of electrons in anisotropic systems depends on the angle of velocity with $x$ axis ($\Phi$). The non-equilibrium distribution function can be written up to the linear order of the electric field as follows\cite{PhysRevB.58.7151}
\begin{equation}
f-f^0=e \left(\frac{\partial f^0}{\partial
	\ve}\right) v_\kv \mathbf E\cdot\boldsymbol\tau(\Phi),
\label{eq:relax-time-generalized}
\end{equation}
where $\boldsymbol\tau(\Phi)$ is the relaxation time vector. In order to obtain the relaxation time vector we define the coefficients $a(\phi)$ and $b(\phi)$ as relaxation times along $x$ and $y$ axes, and write Eq. (\ref{eq:relax-time-generalized}) in the form of Eq. (\ref{f-single-band}) for each branches. In order to use the relation (\ref{f-single-band}) in the Boltzmann equation, we need to write $\Phi_\nu$ (the angle of the velocity of electrons in the branch $\nu$ with $x$ axis) in terms of $\phi$, the polar angle of $\kv$-vector.
In the systems with monotonic band structure electron's velocity ${\bf v}$ is always parallel to the $\kv$-vector and $\Phi_\nu=\phi$. 
But in 2DRSs with strong Rashba SOC, below the BCP electrons' velocity and wave vector are not always in the same direction and depending on the branchband, ${\bf v}$ is parallel or anti-parallel to the $\kv$. By considering the relation $\Phi_{\nu}=\phi+(1+(-1)^\nu)\pi/2$, according to the angular dependence of the functions $a_{\nu}(\Phi_\nu)$ and $b_{\nu}(\Phi_\nu)$ on $\Phi_\nu$ ($a_{\nu}(\Phi)$ and $b_{\nu}(\Phi)$ respectively depend on cosine and sine functions), the relations in Eqs. (\ref{ac12}) and (\ref{bs12}) reduce to the following equations 
\begin{equation}
\cos\phi=\bar{w}_{\nu}(\phi)a_{\nu}(\phi)-\sum_{\nu^{'}}\int d\phi^{'}w_{\nu,\nu^{'}}(\phi,\phi^{'})(-1)^{\nu-\nu^{'}}a_{\nu^{'}}(\phi^{'}),
\label{eq:aaa}
\end{equation}
\begin{equation}
\sin\phi=\bar{w}_{\nu}(\phi)b_{\nu}(\phi)-\sum_{\nu^{'}}\int d\phi^{'}w_{\nu,\nu^{'}}(\phi,\phi^{'})(-1)^{\nu-\nu^{'}}b_{\nu^{'}}(\phi^{'}).
\label{eq:bbb}
\end{equation}
The same as the two-band scattering case, by employing the Fourier series of $a_\nu (\phi)$ and $b_\nu (\phi)$ and performing some straightforward calculations, we reach to the relaxation times in the branches 1 and 2, and finally obtain the conductivity and the AMR of the system.

\end{widetext}


\bibliography{Zarezad-2DRS}

\end{document}